\documentclass[aps,prb,twocolumn]{revtex4-2}

\newcommand{\be}{\begin{equation}}
\newcommand{\ee}{\end{equation}}
\newcommand{\w}{\omega}
\newcommand{\wLR}{\omega_{\rm LR}}
\newcommand{\wq}{\omega_{10}}

\newcommand{\e}{\epsilon}
\newcommand{\qp}{\mathrm{qp}}
\newcommand{\dL}{\Delta_{\rm L}}
\newcommand{\dR}{\Delta_{\rm R}}
\newcommand{\DelBar}{\bar\Delta}
\newcommand{\xL}{x_{\rm L}}
\newcommand{\xR}{x_{\rm R}}
\newcommand{\xRp}{x_{\rm R>}}
\newcommand{\xRm}{x_{\rm R<}}
\newcommand{\Bparone}{B_{\mathrm{\parallel}, 1}}
\newcommand{\Bpartwo}{B_{\mathrm{\parallel}, 2}}
\usepackage{hyperref} 
\usepackage{graphicx}
\usepackage{braket}
\usepackage{amsmath, amsthm, amssymb}
\usepackage{epstopdf}
\usepackage{epsfig}
\usepackage{epsf}
\usepackage{array}
\usepackage{mathtools}
\usepackage{circuitikz}
\usepackage{dcolumn}
\usepackage[tight]{subfigure}
\usepackage[normalem]{ulem}
\usepackage{verbatim}
\usepackage{inputenc}
\usepackage{color}
\usepackage{bm} 
\usepackage{xspace}
\usepackage{mathbbol}
\usepackage{cleveref}
\usepackage{mathcomp}
\usepackage{soul}
\usepackage{upgreek}
\usepackage{appendix}
\usepackage{graphicx}
\usepackage{dcolumn}
\usepackage{bm}
\usepackage{xr}
\makeatletter

\begin{document}

\title{Nonequilibrium regimes for quasiparticles in superconducting qubits with gap-asymmetric junctions} 

\author{Giampiero Marchegiani}
\email{giampiero.marchegiani@tii.ae}
\affiliation{Quantum Research Center, Technology Innovation Institute, Abu Dhabi 9639, UAE}

\author{Gianluigi Catelani}
\affiliation{JARA Institute for Quantum Information (PGI-11),Forschungszentrum J\"ulich, 52425 J\"ulich, Germany}
\affiliation{Quantum Research Center, Technology Innovation Institute, Abu Dhabi 9639, UAE}

\date{\today}

\setlength{\arraycolsep}{2pt}

\begin{abstract} 
Superconducting qubits holds promise for quantum computing, but their operation is challenged by various sources of noise, including excitations known as quasiparticles. Qubits with gap asymmetry larger than their transition energy are less susceptible to quasiparticle decoherence as the quasiparticles are mostly trapped in the low-gap side of the junction. Because of this trapping, the gap asymmetry can contribute to maintaining the quasiparticles out of equilibrium. Here we address the temperature dependence of the quasiparticle densities in the two sides of the junction. We show that four qualitatively different regimes are possible with increasing temperature: i) nonequilibrium, ii) local quasiequilibrium, iii) global quasiequilibrium, and iv) full equilibrium. We identify shortcomings in assuming global quasiequilibrium when interpreting experimental data, highlighting how measurements in the presence of magnetic field can aid the accurate determination of the junction parameters, and hence the identification of the nonequilibrium regimes. 
\end{abstract}

\date{\today}


\maketitle
\section{Introduction}

Superconducting qubits are intensively investigated for quantum computing purposes~\cite{KjaergaardAnnualReview}. Despite  tremendous improvement in performances over the last two decades, still more work is needed for
the precise determination and the mitigation of qubit decoherence processes~\cite{SiddiqiReview}, which are necessary to realize a fault-tolerant large-scale quantum processor. Bogoliubov quasiparticles, the fundamental excitations in superconductors, couple to the qubit when tunneling across Josephson junctions, leading to decoherence~\cite{CatelaniPRL106,CatelaniPRB84,CatelaniSciPostReview} and limiting the fidelity of some two-qubit gates~\cite{Papic2024}. Moreover, quasiparticle bursts can cause correlated errors among distant qubits~\cite{Wilen2021,McEwen2021,li2024direct,harrington2024synchronous}, which can require increased overhead to implement quantum error correction~\cite{CREPRL129}. Quasiparticle mitigation can be achieved via gap-engineering techniques~\cite{Aumentado2004,SunPRL108,RiwarPRB94}, which can keep quasiparticles away from the qubit's junctions, confining them in regions with lower superconducting gap. Recent research highlighted intrinsic gap-engineering effects in gap-asymmetric Josephson junctions~\cite{Diamond22,MarchegianiQP}: in standard nanofabricated Al-AlOx-Al Josephson junctions, the top layer has a larger thickness than the bottom one to ensure the continuity of the film, resulting in a gap asymmetry $\delta\Delta$ due to the strong thickness modulation of the Al superconducting gap below 100~nm. 
In typical devices, the gap-asymmetry frequency $\omega_{\rm LR}/2\uppi=\delta\Delta/(2\uppi\hbar)$ ($\hbar$ is the reduced Planck's constant, which we set to unity from now on) can be of the order of a few GHz, and is thus comparable, or even exceeding, the qubit frequency $\wq/2\uppi$. In fact, qubits fabricated with $\wLR>\wq$ are protected from quasiparticle tunneling, as quasiparticles are trapped in the low-gap electrode~\cite{MarchegianiQP}; the advantage of this design was experimentally confirmed in a 3D transmon~\cite{connolly2023coexistence}, and it has also been proven effective in suppressing correlated errors in multi-qubit devices~\cite{mcewen2024resisting}.  

From a complementary point of view, the accurate characterization of superconducting qubits has enabled detailed studies of physical effects that are, in general, more difficult to investigate in other setups: evidence for quasiparticle interference has been reported in $T_1$ measurements of a fluxonium~\cite{Pop2014}; trapping of quasiparticle by one or more vortices has been measured in a transmon~\cite{Wang2014}; higher harmonics in the current-phase relation have been shown to have an influence on the energy level and charge dispersion in transmons larger than naively expected for tunnel junctions~\cite{Willsch24}. In contrast, while the gap asymmetry itself can be directly extracted from transport measurements~\cite{Barone}, it is not so easily estimated in qubits; still, its presence leads to a peak in transition rates of a SQUID transmon~\cite{Diamond22}, or to thermal activation of the rates when $k_{\rm B} T\sim \delta\Delta$~\cite{connolly2023coexistence} (where $T$ is the temperature and $k_{\rm B}$ is the Boltzmann constant). In the latter work, the thermal activation behavior is considered as a signature of the coexistence between an excess quasiparticle number and Fermi-like energy distribution of the quasiparticles. Here we extend the low temperature ($k_{\rm B} T \ll \wLR, \, \wq$) modeling of qubit-quasiparticle interaction~\cite{MarchegianiQP} to cover the intermediate temperature range up to $k_{\rm B} T \lesssim \wLR, \, \wq$. With our approach, we can address the temperature dependence of the quasiparticle density, and we show that different nonequilibrium regimes can be possible, depending on the gap asymmetry and the quasiparticle generation rates. We discuss possible limitations of what we call global quasiequilibrium modeling, which may result in an inaccurate estimate of the qubit's parameters, and explore possible measurement schemes to overcome them, in particular exploiting the Fraunhofer effect~\cite{Krause2022,Krause24} in the presence of a magnetic field parallel to the junction plane.
\section{Results}
\subsection{Model} \label{sec:ansatzQP}
\begin{figure}
  \begin{center}
    \includegraphics[]{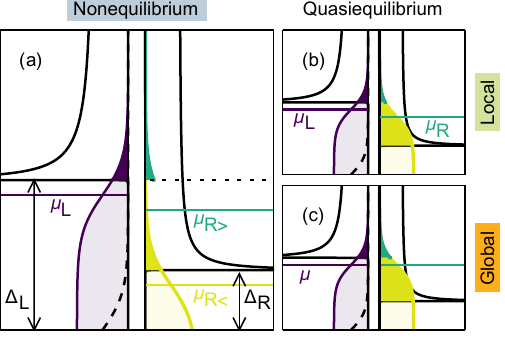}
  \end{center}
  \caption{\textbf{Schematic of different nonequilibrium regimes.} In each panel we show the density of states
    and quasiparticle distribution functions in the two superconducting electrodes forming the Josephson junction of a transmon qubit. 
    The two electrodes are characterized by a gapped excitation spectrum with asymmetric gaps, $\dL>\dR$. We consider Fermi distributions with characteristic temperature $T$ corresponding to the phonon bath temperature. The excess quasiparticle density in the high-gap electrode is accounted for through a chemical potential $\mu_{\rm L}$ (at thermal equilibrium, the chemical potential coincides with the Fermi energy, $\mu_{\rm L}=0$). In the right electrode, we distinguish between excitations with energy smaller or larger than $\dL$: correspondingly, we identify two different values for the chemical potential, $\mu_{\rm R>}$ and $\mu_{\rm R<}$. The different regimes are: (a) full nonequilibrium (b) local quasiequilibrum ($\mu_{\rm R>}=\mu_{\rm R<}\neq\mu_{\rm L}$), and (c) global quasiequilibrium ($\mu_{\rm R>}=\mu_{\rm R<}=\mu_{\rm L}$).}
  \label{fig:Fig1}
\end{figure}

For typical qubits based on Al-AlOx-Al Josephson junctions, the qubit transition rates can usually be measured from the base temperature of the dilution refrigerator, $T\sim 10\,$mK, up to roughly $250\,$mK; at this temperature, nonequilibrium effects are expected to be negligible and the qubit's lifetime is usually too short to be of practical interest. In this range the condition $k_{\rm B} T\ll \Delta$ is well satisfied for aluminum thin film, characterized by a critical temperature $T_{\rm C}=\Delta/(1.764 k_{\rm B})\sim 1.6~$K, [here we use the standard relation between the critical temperature and the zero-temperature gap $\Delta$ in the Bardeen-Cooper-Schrieffer (BCS) theory~\cite{tinkham}]. Thus, we can safely disregard gap suppression effects due to temperature. We consider a superconducting qubit comprising a single Josephson junction (JJ), \textit{i.e.}, two superconducting films, denoted with left (L) and right (R), separated by a thin, nanometer-scale insulating barrier. Unless explicitly specified, we set $k_{\rm B} = 1$ hereinafter. 
We aim to describe the evolution with temperature of the qubit-quasiparticle system, incorporating nonequilibrium effects. Specifically, we consider the following ansatz for the quasiparticle energy distributions in the left ($f_{\rm L}$) and right ($f_{\rm R}$) electrodes forming the Josephson junction, 
\begin{align}
f_{\rm L}(\e)&= f_0(\e-\mu_{\rm L})\,,
    \label{eq:fLAns}
    \\
f_{\rm R}(\e)&=
f_0(\e-\mu_{\rm R<})\theta(\dL-\e)
    +f_0(\e-\mu_{\rm R>})
    \theta(\e-\dL)\,,
    \label{eq:fRAns}
\end{align}
where $\e$ is the quasiparticle energy measured with respect to the Fermi level, $\theta(\epsilon)$ is the Heaviside step function, $f_0(\e)=[\exp(\e/T)+1]^{-1}$ is the Fermi-Dirac distribution, and $T$ is the quasiparticle temperature. We take the latter to coincide with the phonon temperature of the substrate~\cite{connolly2023coexistence}-- we discuss limitations to this assumption in Sec.~\ref{sec:conclusions}. 
We account for the possible nonequilibrium nature of the quasiparticles via the effective chemical potentials $\mu_{\rm L}$, $\mu_{\rm R<}$, and $\mu_{\rm R>}$ -- the equilibrium case being $\mu_{\rm L}=\mu_{\rm R<}=\mu_{\rm R>}=0$. In our approach, the chemical potentials will be an output of our model and are determined by the interplay between the different physical processes involving the quasiparticles (detailed in Sec.~\ref{sec:QPdyn}). This approach generalizes that pioneered by Owen and Scalapino~\cite{OwenScalapinoPRL28}, which has already been used in modeling superconducting qubits~\cite{PalmerPRB76,CatelaniPRB89,Ansari_2015}. In Fig.~\ref{fig:Fig1}, we give a schematic representation of our ansatz. Specifically, we display the quasiparticle density of states (DoS) in the two superconducting electrodes of the JJ, characterized by the energy gaps $\dL$ and $\dR$; with no loss of generality, we assume $\dL\geq\dR$ hereinafter. The quasiparticle distributions are represented via filled curves: the darker regions correspond to occupied states, while the lighter fillings characterize unavailable states. The occupation of states with energy larger than the chemical potential is exponentially suppressed with the ratio $(\e-\mu)/T$, as per our ansatz [cf. Eqs.~\eqref{eq:fLAns} and \eqref{eq:fRAns}]. We remark that the three chemical potentials are, in general, different; in Fig.~\ref{fig:Fig1}a, we consider a specific order, i.e., $\mu_{\rm L}>\mu_{\rm R>}>\mu_{\rm R<}$, which can occur at low temperature (see Sec.~\ref{sec:chemPot}). In Fig.~\ref{fig:Fig1}b the low-gap electrode is characterized by a single chemical potential, $\mu_{\rm R>}=\mu_{\rm R<}\neq\mu_{\rm L}$ (local quasiequilibrium), while in Fig.~\ref{fig:Fig1}c the three chemical potentials are the same, a situation which we call global quasiequilibrium. For reference, we show with a dashed line the quasiparticle distribution in the left electrode at equilibrium, $\mu_{\rm L}=0$.

\subsection{Rate equations} 
\label{sec:QPdyn}
\begin{figure}
  \begin{center}
    \includegraphics[]{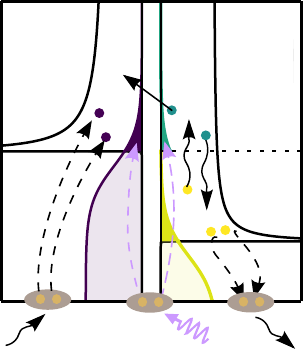}
  \end{center}
  \caption{\textbf{Processes governing the quasiparticle dynamics.} The upward diagonal arrow represents a qubit relaxation event in which a quasiparticle tunnels from the right to the left electrode. Wavy arrows denote phonon-mediated processes: the downward arrow shows a quasiparticle relaxation process accompanied by phonon emission, while the upward one excitation by phonon absorption. Phonons can also break Cooper pairs (see bottom of left electrode), or be emitted in a recombination event (see right electrode). Pair-breaking photons lead to generation of one quasiparticle in each electrode at the same time.}
  \label{fig:Fig1point5}
\end{figure}

Since in this work we investigate the nonequilibrium steady state of the quasiparticle-qubit system in the absence of any coherent drive, we can simply focus on the diagonal part of the density matrix, whose dynamics is governed by rate equations. When dephasing effects need to be considered, a more general approach is required, see e.g. ~\cite{CatelaniPRB86} and \cite{CatelaniPRB89}. 
We consider low temperatures compared to the gap; more precisely, we assume $T\ll \Delta-\mu$, so the quasiparticle occupation probability is small, $f_{\rm L},\,f_{\rm R} \ll 1$. Thus, we can equivalently consider the quasiparticle densities $\xL$, $\xRm$, and $\xRp$~\cite{MarchegianiQP}, which are in one-to-one correspondence with the chemical potentials (the relation is given in Appendix~\ref{app:XqpandChemPot}). This choice is convenient since, with our assumption, the rates of transition between qubit states induced by quasiparticles are proportional to their densities, and it eases comparison with previous literature (see Ref.~\cite{CatelaniSciPostReview} and references within).
Following Ref.~\cite{MarchegianiQP}, we define $\xRp\,(\xRm)$ as the quasiparticle density in the low-gap electrode with energy larger (smaller) than $\dL$~\cite{MarchegianiQP}. 

The rate equations for the occupation probability $p_i$ of the qubit logical state $i$ (with $\bar{i}=1$ for $i=0$ and \textit{vice versa}) read
\be
\dot{p}_i=-(\Gamma_{i\bar i}^{\rm eo}+\Gamma_{i\bar i}^{\rm ee})p_i
+(\Gamma_{\bar i i}^{\rm eo}+\Gamma_{\bar i i}^{\rm ee})p_{\bar i} \, ,
\label{eq:Qubit_rate}
\ee
where the dot denotes the time derivative and we explicitly distinguish between rates preserving and changing the parity (${\rm e}=\,$even and ${\rm o}=\,$odd); here the parity is that of the number of quasiparticles that have tunnelled across the junction~\cite{CatelaniSciPostReview,CatelaniPRB89} and we assume the rates to be symmetric when exchanging parities, ${\rm e} \leftrightarrow {\rm o}$, see Appendix~\ref{app:qubitPopulation}. Parity-preserving rates are of non-quasiparticle origin; in this work we assume that such transitions are caused by a thermal bath and hence, according to the detailed balance principle, we have $\Gamma_{01}^{\rm ee}=\exp(-\w_{10}/T)\Gamma_{10}^{\rm ee}$. In contrast, rates modifying the parity are related to quasiparticle events; they can be expressed as $\Gamma_{ii'}^{\rm eo}=\Gamma^{\rm ph}_{ii'}+\sum_\alpha \tilde\Gamma_{ii'}^{\alpha} x_\alpha$, where $\Gamma^{\rm ph}_{ii'}$ accounts for photon-assisted transition~\cite{PRL123Photon}; these transitions are caused by pair-breaking photons and contribute to quasiparticle generation (see Sec.~\ref{sec:QPgen}). 
In the sum, $\alpha={\rm L,\,R>,\, R<}$ denotes the initial location of the quasiparticle tunneling through the junction.

The rate equations for $\xL,\xRp,\xRm$ were introduced and discussed in detail in the low-temperature limit $T\ll \w_{10},\,\wLR,\, |\w_{10}-\wLR|$~\cite{MarchegianiQP}; here, we present them in a slightly generalized form to incorporate physical processes that become important for temperatures comparable to the other three energy scales:
\begin{align}
\dot x_{\rm L} &= g^{\rm L}-r^{\rm L}\xL^2
-\delta[(\bar\Gamma_{00}^{\rm L}+\bar\Gamma_{01}^{\rm L})p_0
+(\bar\Gamma_{11}^{\rm L}+\bar{\Gamma}^{\rm L}_{10})p_1]\xL \nonumber\\
&+\delta\bar{\Gamma}^{\rm R<}_{10}p_1 \xRm \nonumber\\
&+\delta[(\bar\Gamma_{00}^{\rm R>}+\bar{\Gamma}^{\rm R>}_{01})p_0+(\bar\Gamma_{11}^{\rm R>}+\bar{\Gamma}^{\rm R>}_{10})p_1] \xRp \, ,
\label{eq:dotxL}\\
\dot x_{\rm R>}&=g^{\rm R>}-r^{\rm R>} \xRp^2 -r^{<>}\xRm\xRp \nonumber\\
&- [(\bar\Gamma_{00}^{\rm R>}+\bar{\Gamma}^{\rm R>}_{01})p_0+(\bar\Gamma_{11}^{\rm R>}
+\bar\Gamma_{10}^{\rm R>})p_1]\xRp \nonumber\\
&+[\bar\Gamma_{00}^{\rm L}p_0+(\bar\Gamma_{11}^{\rm L} + \bar\Gamma_{10}^{\rm L})p_1]\xL+\xi \bar\Gamma_{01}^{\rm L}p_0\xL  \nonumber\\
& -\tau_{\rm R}^{-1}\xRp  +\tau_{\rm E}^{-1}\xRm \, ,
\label{eq:dotxR>} \\
\dot
x_{R<}&=g^{R<}-r^{R<} \xRm^2 -r^{<>}\xRm\xRp -\bar{\Gamma}^{R<}_{10}p_1 \xRm 
\nonumber\\  
&  +\left(1-\xi\right)\bar\Gamma_{01}^L p_0 \xL  +\tau_R^{-1}\xRp -\tau_E^{-1}\xRm \, ,
\label{eq:dotxR<}
\end{align}
where the gap ratio $\delta=\dR/\dL$ in Eq.~\eqref{eq:dotxL} is related to our normalization choice of the quasiparticle densities (see Appendix~\ref{app:XqpandChemPot}), which differs from the one made elsewhere~\cite{connolly2023coexistence}. 
The quasiparticles processes which we discuss next are schematically summarized in Fig.~\ref{fig:Fig1point5}. In the right-hand side (RHS) of Eqs.~\eqref{eq:dotxL}-\eqref{eq:dotxR<}, terms associated with a positive sign increase the quasiparticle density, while terms with a negative sign reduce the density. The total quasiparticle number in the device is $N_{\rm qp}=2\nu_0 \dL\mathcal V[\xL + \delta(\xRm + \xRp)]$, where we assumed for simplicity the single-spin densities of states at the Fermi level in each electrode $\nu^{\rm L}=\nu^{\rm R}=\nu_0$ as well as the electrodes' volumes $\mathcal V_{\rm L}=\mathcal V_{\rm R}=\mathcal V$ to be the same. The total quasiparticle number can be increased via quasiparticle generation, see terms $g^{\rm L}$, $g^{\rm R<}$, and $g^{\rm R>}$ (see discussion in Sec.~\ref{sec:QPgen}), 
or reduced via quasiparticle recombination. The latter mechanism destroys two quasiparticles either in the left [term proportional to $r^{\rm L}$ in the RHS of Eq.~\eqref{eq:dotxL}]  or in the right [terms proportional to $r^{<>}$, $r^{\rm R>}$ and $r^{\rm R<}$ in the RHS of Eqs.~\eqref{eq:dotxR>} and \eqref{eq:dotxR<}] electrode, with consequent phonon emission.  Terms linear in the densities do not change the total quasiparticle number; rather, they lead to a redistribution of these excitations. In the low-gap electrode, electron-phonon scattering can lead to quasiparticle excitation and relaxation [terms $\tau_{\rm E}^{-1}\xRm$ and $\tau_{\rm R}^{-1}\xRp$ in the RHS of Eqs.~\eqref{eq:dotxR>} and \eqref{eq:dotxR<}, respectively]. The remaining terms are associated with quasiparticle tunneling: these processes always change the qubit's parity (${\rm e}\to {\rm o}$ or ${\rm o}\to {\rm e}$) and can lead to a transition between initial ($i$) and final ($f$) qubit logical states. In the rate equations, they appear in a specific combination of the general form $ \bar{\Gamma}_{if}^\alpha p_i x_\alpha$, where the quantity $\bar{\Gamma}_{if}^\alpha$ 
is the tunneling rate for a single quasiparticle (a different notation is used in other works~\cite{Diamond22,connolly2023coexistence}). 
These rates correspond to the tilde rates in $\Gamma_{\bar i i}^{\rm eo}$ [see Eq.~\eqref{eq:Qubit_rate} and text that follows] divided by the Cooper pair number in the low-gap electrode, i.e.,  $\bar{\Gamma}_{if}^\alpha=\tilde{\Gamma}_{if}^\alpha/(2\nu_0\dR\mathcal V)$~\cite{MarchegianiQP}.
The rates $\bar{\Gamma}_{ii}^{\rm R<}$, $\bar{\Gamma}_{01}^{\rm R<}$ are exactly zero due to our ansatz in Eqs.~\eqref{eq:fLAns} and ~\eqref{eq:fRAns}. The $p_i$ factors in Eqs.~\eqref{eq:dotxL}-\eqref{eq:dotxR<} and the dependence of the rates entering in Eq.~\eqref{eq:Qubit_rate} on the quasiparticle densities couple the dynamics of the qubit and the quasiparticles. Below, we solve these equations in the steady state as a function of quasiparticle temperature for typical experimental parameters, and we identify various nonequilibrium regimes.

Explicit expressions for the rates and their temperature dependence in terms of qubit parameters in the case of a transmon are given in Appendix~\ref{app:transitionRatesSingleJunction}. 
Here we comment on the relation between the above rates equations with those in the low-temperature limit~\cite{MarchegianiQP}. The dimensionless factor $0\le\xi\le 1$ denotes the fraction of quasiparticles tunneling from the left to the right electrode with final quasiparticle energy larger than $\dL$ in a qubit excitation process (see Sec.~\ref{app:assignmentGamma01L}), and it is, in general, a function of $T$, $\wLR$, and $\wq$; at low temperature $T\ll |\wq - \wLR|$, $\xi$ is exponentially suppressed as ${\rm e}^{-\min\{\wLR,\w_{10}\}/T}$.
The quasiparticle excitation rate $\tau^{-1}_{\rm E}$ due to thermal phonons is exponentially small in $\wLR/T$ (see Appendix~\ref{app:qp-phonon}) and was thus disregarded in the low-temperature modelling~\cite{MarchegianiQP}. Finally, the rate $\bar{\Gamma}_{01}^{\rm R>}\propto {\rm e}^{-\w_{10}/T}$ is also exponentially suppressed (see Appendix~\ref{app:transitionRatesSingleJunction}). Hence, in the low-temperature limit $T\ll \w_{10}, \, \wLR,\, |\wLR-\w_{10}|$, the Eqs.~\eqref{eq:dotxL}-\eqref{eq:dotxR<} reduce to those in the literature~\cite{MarchegianiQP}. The generalization given here, as we show below, enables us to study the crossover to full equilibrium.

\subsection{Quasiparticle generation mechanisms}
\label{sec:QPgen}
Quasiparticles are generated by pair-breaking processes due to the interaction of the electrons in the superconductors forming the junction with phonons in the electrodes and with external electromagnetic radiation~\cite{Catelani_Pekola_2022}. Thus, we express the generation rate for the quasiparticle density $x_\alpha$ (with $\alpha=\{\rm L,R<,R>\}$) as the sum of PHotons and PhoNons contributions
\begin{equation}
\label{eq:generationGeneral}
g^\alpha=g_\alpha^{\rm ph}+g_\alpha^{\rm pn} \,.
\end{equation}
Nonequilibrium quasiparticle effects in superconducting qubits and resonators due to the interplay between phonons and photons have previously been investigated in Refs.~\cite{Basko,Fischer2023}, and more recently including photon pair-breaking processes~\cite{Fischer2024}.
In this work, we assume that the photon absorption mainly occurs at the junction due to its high impedance with respect to the rest of the circuit~\cite{LiuPRL132}. Absorption of photons with energy $\omega_\nu$ larger than the gap sum $\dL+\dR$ causes photon-assisted tunneling processes~\cite{PRL123Photon}, leading both to quasiparticle generation and parity switching. The generation rate by pair-breaking photons in electrode $\alpha=\{\rm L,R\}$ is obtained by dividing the total photon-assisted rate $\Gamma^{\rm ph}=\sum_{i,j} p_i \Gamma^{\rm ph}_{ij}$ by the Cooper pair number in $\alpha$, i.e., $g_{\rm R}^{\rm ph}=\Gamma^{\rm ph}/(2\nu_0\dR\mathcal V)=g_{\rm L}^{\rm ph}/\delta$~(the $\delta$ at the denominator is related again to our normalization choice~\cite{MarchegianiQP}); the separation of $g^{\rm ph}_{\rm R}$ into generation rates for quasiparticles with energy smaller ($g^{\rm ph}_{\rm R<}$) and larger ($g^{\rm ph}_{\rm R>}$) than $\dL$ and expressions for $\Gamma^{\rm ph}_{ij}$ are discussed in Appendix~\ref{app:PAT}.

Phonon absorption mainly occurs in the bulk, and lead to quasiparticle generation in electrode $\alpha=\{\rm L,R\}$ above the energy threshold $2\Delta_\alpha$. 
Recent works have shown that pair-breaking phonons generated by absorption of background radiation can threaten the operation of superconducting qubits~\cite{GordonAPL120,Vepslinen2020,Cardani2021,Wilen2021}. Such radiation can lead to quasiparticle bursts and so a temporary increase in the quasiparticle density~\cite{PRL121Res,Cardani2021,McEwen2021}; the precise description of this time-resolved dynamics~\cite{YeltonPRB110} goes beyond the scope of this manuscript. Hence, in this work, we disregard quasiparticle generation due to nonequilibrium phonons, which can also be suppressed using phonon traps~\cite{APLphonon,IaiaNatComm}.
Conversely, since we are primarily interested in discussing the evolution with temperature of the quasiparticle-qubit system, we include the contribution of thermal phonons, and express the generation rates by phonons as $g_{\rm L}^{\rm pn}=2\uppi r^{\rm L} T {\rm e}^{-2\dL/T}/\dL$, $g_{\rm R<}^{\rm pn}=2\uppi r^{\rm R<} T {\rm e}^{-2\dR/T}{\rm erf}[\sqrt{\wLR/T}]/\dR$, $g_{\rm R>}^{\rm pn}=2\uppi r^{\rm R<} T {\rm e}^{-2\dR/T}{\rm erfc}[\sqrt{\wLR/T}]/\dR$; these expressions are valid for $T\ll\dL,\dR$, see Appendix~\ref{app:qp-phonon}.

\subsection{Chemical potentials vs temperature}
\label{sec:chemPot}
\begin{figure}
  \begin{center}    \includegraphics[]{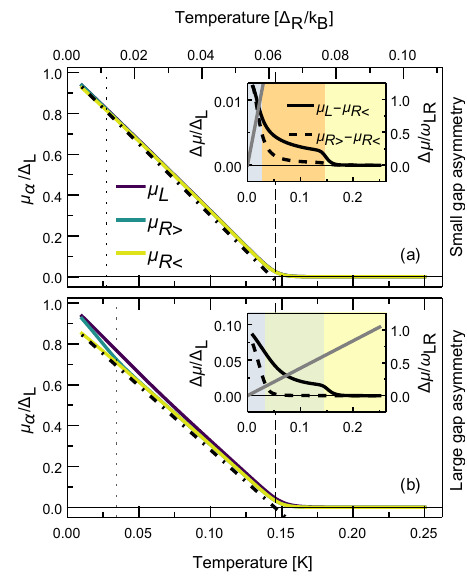}
  \end{center}
  \caption{\textbf{Chemical potentials vs temperature.}  Temperature dependence of the chemical potentials $\mu_{\rm L}$, $\mu_{\rm R>}$, and $\mu_{\rm R<}$ (cf. Fig.~\ref{fig:Fig1}a) in a single-junction transmon (charging energy $E_{\rm C}$ and Josephson energy $E_{\rm J}$): (a) small ($\wLR/2\uppi=0.5$~GHz) or (b) large ($\wLR/2\uppi=5$~GHz) gap asymmetry. The dot-dashed lines represent constant quasiparticle density, as discussed in the ``Chemical potentials vs temperature'' subsection. Insets: chemical potential differences ($\Delta\mu$) vs temperature. The grey solid lines show the thermal energy for comparison. Note the two scales on the left and the right vertical-axis are normalized with the left electrode gap $\dL$ and the gap asymmetry $\wLR$, respectively. Colored regions identify the different regimes: nonequilibrium (aquamarine, left), global quasiequilibrium (orange, middle) or local quasiequilibrium (light green, middle), and equilibrium (yellow, right) -- see also the highlighted labels in Fig.~\ref{fig:Fig1}. The temperature values separating the regimes are identified with dotted and dashed vertical lines in the main panels. Parameters: $\dR/h=49$~GHz, $\wq/2\uppi=5.5~$GHz, $E_{\rm J}/h=14.5$~GHz, $E_{\rm C}/h=290$~MHz, $\omega_{\nu}/2\uppi=119$~GHz, $\Gamma_{00}^{\rm ph}=300$~Hz, $\nu_0=0.73\times 10^{47}\,{\rm J}^{-1}{\rm m}^{-3}$, $\mathcal{V}=506\times240\times0.028~\mu{\rm m}^3$, $\Gamma_{10}^{\rm ee}=100$~kHz, $r^{\rm L}=r^{\rm R<}=6.25$~MHz. 
  }
  \label{fig:Fig3}
\end{figure}

Having introduced our models for the coupled quasiparticle-qubit system, we now move to considering its steady state as function of temperature $T$. In this case, the time derivatives on the left-hand sides of Eqs.~\eqref{eq:Qubit_rate}-\eqref{eq:dotxR<} vanish, and those equations reduce to algebraic ones. Still, the system being nonlinear, analytical solutions can only be found approximately in certain limits, as we discuss in Sec.~\ref{app:QPDensApprox}. 
 Here, we numerically find the roots of the system of coupled equations; subsequently, we calculate the chemical potentials $\mu_{\rm L}$, $\mu_{\rm R>}$, and $\mu_{\rm R<}$, using the computed steady-state values of the quasiparticle densities (cf. Appendix~\ref{app:XqpandChemPot}). In Fig.~\ref{fig:Fig3}, we display our results for the cases of small $\wLR/(2\uppi)= 0.5$~GHz (panel a) and large gap asymmetry $\wLR/(2\uppi)= 5$~GHz (panel b) in relation to thermal energy  $T/(2\uppi)= 0.2~$GHz at the base temperature of a typical dilution refrigerator, $T\approx 10$~mK. In the calculation, we set the photon-assisted parity-switching rate in the qubit ground state to $\Gamma_{00}^{\rm ph}=300~$Hz, comparable to the values reported by several experiments~\cite{SerniakPRApp,Riste13,ChristensenPRB100,kurter2022quasiparticle,Krause24}, but few orders of magnitudes larger than that in state-of-the-art devices optimized to suppress such contributions~\cite{IaiaNatComm,Fei2022,connolly2023coexistence,YeltonPRB110}. Other parameters are specified in the caption. In both cases, all three chemical potentials are the largest at the lowest temperature considered and decrease monotonically (approximately linearly), reaching zero, i.e., full equilibrium, around $T\gtrsim 150~$mK for our parameter choice. This behavior arises from the competition between photon-assisted pair-breaking, which is temperature-independent, and pair-breaking by thermal phonons, whose rate increases exponentially with temperature. Since $\dR<\dL$, a crossover temperature $\bar{T}$ can be defined by comparing photon generation to thermal phonon generation in the right electrode, $g_{\rm R>}^{\rm pn}+g_{\rm R<}^{\rm pn}=g_{\rm R}^{\rm ph}$, leading to the expression~\cite{Fischer2024}
\begin{equation}
\label{eq:crossoverTemp}
\bar{T} = \frac{2\dR}{W\left( 4\uppi r^{\rm R<}/g_{\rm R}^{\rm ph} \right)}   \, ,
\end{equation}
where $W(z)$ is the Lambert W (or product-log) function. For our parameters, this formula provides an estimate of the crossover within a few percent of the numerical results in Fig.~\ref{fig:Fig3}, as shown by the dashed vertical lines in the figure.

The approximately linear decrease of the chemical potentials with temperature signals a roughly constant quasiparticle density for temperatures below $\bar{T}$, since 
$\mu_\alpha\approx \Delta_\alpha- T\, \log(1/x_\alpha)$ (with a slight abuse of notation, we mean here $\Delta_{\rm R<}\equiv\Delta_{\rm R>}\equiv\dR$); this linear decrease is displayed for $\alpha={\rm R}<$ in Fig.~\ref{fig:Fig3}a and b with dot-dashed lines. Below we comment on the deviation from approximately linear behavior for $\mu_{\rm R>}$ at large gap asymmetry; in fact, the relation between the chemical potentials depends on the gap asymmetry, reflecting different energy distributions in the two cases. For a small $\wLR$, the nonequilibrium quasiparticles are approximately equally distributed in the two electrodes, $\xL\approx \xRp+\xRm\approx \sqrt{g_{\rm R}^{\rm ph}/r^L}$. At larger $\wLR$, quasiparticles accumulate in the low-gap electrode, $\xRm\gg\xL\gg\xRp$, due to fast relaxation compared to tunneling~\cite{MarchegianiQP}. In both cases, the chemical potential differences (see insets of Fig.~\ref{fig:Fig3}a and b) are largest at the lowest temperature considered, where $\mu_{\rm L}-\mu_{\rm R>}\sim \mu_{\rm R>}-\mu_{\rm R<}\sim \wLR$ (slightly greater or smaller than $\wLR$ for small and large gap asymmetry, respectively; see also Appendix~\ref{app:chemDiff}), and drop as quasiparticle-phonon scattering rates increase with temperature. Hence, at low asymmetry, the chemical potential differences become smaller than
the thermal energy (solid grey line in Fig.~\ref{fig:Fig3}a) already at low temperature, and the system is approximately in global quasiequilibrum (see also Sec.~\ref{sec:impactRates}). In contrast, the chemical potential differences can remain above the thermal energy up to higher temperatures
in the gap-asymmetric case. 
This makes it possible to clearly identify different regimes: for $T\lesssim 30~$mK the three chemical potentials are unequal $\mu_{\rm L}>\mu_{\rm R>}>\mu_{\rm R<}$ (the nonequilibrium regime of Fig.~\ref{fig:Fig1point5}a); for $30\,$mK~$\lesssim T\lesssim \bar{T} \simeq 150\,$mK
the low-gap electrode can be characterized using a single chemical potential, $\mu_{\rm R>}=\mu_{\rm R<}\neq \mu_{\rm L}$ (local quasiequilibrium, Fig.~\ref{fig:Fig1}b); finally, above $\bar{T}$ the three chemical potentials fully equilibrate, $\mu_{\rm R>}\approx \mu_{\rm R<}\approx \mu_{\rm L} \ll T$. 
The transition between nonequilibrium and local quasiequilibrium and the corresponding deviation from linearity for $\mu_{\rm R>}$ vs $T$ are due to the interplay between tunneling from the high-gap to the low-gap electrode and the excitation rate due to phonon absorption in the low-gap electrode: at sufficiently low temperatures, quasiparticle excitations is slow compared to tunneling, leading to a nonequilibrium accumulation in the low gap-electrode, $\mu_{\rm R>}>\mu_{\rm R<}$; as the temperature increases, the excitation rate grows exponentially with temperature, becomes faster than tunneling, and we have $\mu_{\rm R>}$ approaching $\mu_{\rm R<}$ (see Appendix~\ref{app:crossoverNELocal}). The dotted vertical lines in Fig.~\ref{fig:Fig3}a and b identify the temperature at which the tunneling from the left to the right electrode at energies $\e\geq\dL$ equates the excitation rate (see Appendix~\ref{app:crossoverNELocal}); below this temperature we have the nonequilibrium regions (aquamarine) highlighted in the insets.
\subsection{Impact on qubit transition rates}
\label{sec:impactRates}
\begin{figure}
   \begin{center}    \includegraphics[]{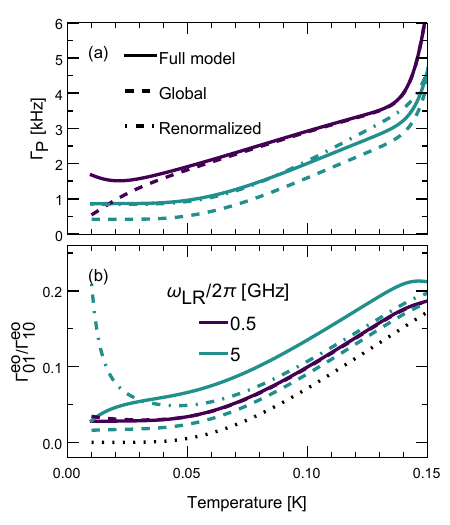}
   \end{center}
   \caption{\textbf{Parity-switching rates vs temperature.}
        (a) Total parity-switching rate and (b) excitation to relaxation ratio vs temperature for small and large gap asymmetry ($\wLR$). Parameters are given in the caption of Fig.~\ref{fig:Fig3}. For the dot-dashed curves, $\Gamma_{00}^{\rm ph}=600$~Hz and $\wLR/(2\uppi)=6$~GHz.
   }
   \label{fig:Fig4}
\end{figure}

The measurement of parity-switching rates is well-established as a probe of quasiparticle effects~\cite{Riste13,CatelaniPRB89,PRL121Exp}, so we consider now whether such measurements can help distinguishing the various nonequilibrium regimes. In Fig.~\ref{fig:Fig4}a we display the parity switching rate $\Gamma_{\rm P}=p_0 (\Gamma_{01}^{\rm eo}+\Gamma_{00}^{\rm eo})+p_1 (\Gamma_{10}^{\rm eo}+\Gamma_{11}^{\rm eo})$ calculated using the same parameters as in Fig.~\ref{fig:Fig3} (solid curves). The parity-switching rate increases monotonically for the large-asymmetry junction, while for the small-asymmetry one there is a nonmonotonic evolution at low temperatures $T\lesssim 25~$mK. This nonmonotonic behaviour stems from the competition between the tunneling rate $\bar\Gamma_{00}^{\rm R>}$ and relaxation in the low-gap electrode ($\tau_{\rm R}^{-1}$), favoring the accumulation of quasiparticles at energies above $\dL$ when temperature becomes smaller than the gap difference (see Appendices~\ref{app:transitionRatesSingleJunction} and \ref{app:QPDensApprox}). This interplay is absent for large gap-asymmetries, since relaxation is fast compared to tunneling $\tau_{\rm R}^{-1}\gg \bar\Gamma_{00}^{\rm R>}$~\cite{MarchegianiQP}, leading to accumulation of quasiparticles at the low-gap energy. For comparison, we also display the parity-switching rates calculated assuming global quasiequilibrium (dashed curves), in which case the ratio between the quasiparticle densities is fixed by the gap asymmetry and the temperature (see last paragraph in Appendix~\ref{app:chemDiff}) and the overall density by the total generation rate, cf. Eq.~\eqref{eq:generationGeneral}. For small gap asymmetry, the global quasiequilibrium modeling reproduces accurately the full nonequilibrium calculation, except at the lowest temperatures; the region of agreement coincides with that where the chemical potential differences are smaller than temperature, see inset in Fig.~\ref{fig:Fig3}a. The low-temperature nonmonotonic feature is missed because by requiring equal chemical potentials the thus constrained model underestimates the number of quasiparticles with energy larger than $\dL$~\cite{MarchegianiQP}, since in fact we have $\mu_{\rm L}\simeq\mu_{\rm R>}>\mu_{\rm R<}$. However, this discrepancy may be difficult to uncover experimentally, as achieving thermalization is progressively harder closer to the base temperature of the cryostat~\cite{Giazotto2006}. For large asymmetry, measurement of $\Gamma_{\rm P}$ also does not permit distinguishing global quasiequilibrium from full or local nonequilibrium: the temperature behaviour is qualitatively similar, and by repeating the calculation with different photon rate and the gap asymmetry (increased by factors of $~2$ and $\sim 1.2$,  respectively, for our parameter choice), the global quasiequilibrium approach can describe reasonably well the results of the full model. Therefore, even though one could fit experimental data for parity lifetime vs temperature assuming global quasiequilibrium, the extracted parameter values would not be accurate. 

The analysis above shows that additional measurements are needed to identify the nonequilibrium regimes. One possibility is to consider the temperature dependence of the ratio between excitation and relaxation rates for parity switching transitions, as displayed in Fig.~\ref{fig:Fig4}b. For this quantity, the difference between the global quasiequilibrium and the full nonequilibrium modeling is hardly noticeable for small gap asymmetry, while more prominent deviations occur for the large-asymmetry junction. Importantly, the deviations become even more significant when the parameters are rescaled as discussed above (cf. Fig.~\ref{fig:Fig4}a) to capture the total parity lifetime; in other words, it is in general not possible to consistently fit data for both $\Gamma_{\rm P}$ and $\Gamma^{\rm eo}_{10}/\Gamma^{\rm eo}_{10}$ under the assumption of global quasiequilibrium. For our specific parameters, the order relation between gap asymmetry and frequency switches from $\wLR<\wq$ to $\wLR>\wq$ after rescaling. In the latter case the rate $\bar{\Gamma}_{10}^{\rm R<}$ is exponentially suppressed with $(\wLR-\wq)/T$ (cf. Appendix~\ref{app:matTra}), resulting in the strong upturn of $\Gamma_{01}^{\rm eo}/\Gamma_{10}^{\rm eo}$ at low temperatures.
For both small and large asymmetry the excitation/relaxation ratio deviates from the detailed balance expectation $\Gamma_{01}^{\rm eo}/\Gamma_{10}^{\rm eo}=\exp(-\wq/T)$ (dotted line) because for photon-assisted transitions the excitation and relaxation rates are typically of the same order, $\Gamma_{01}^{\rm ph}/\Gamma_{10}^{\rm ph}\approx 1$~\cite{PRL123Photon}. In summary, a joint measurement of the parity lifetime and the ratio of the excitation and relaxation rates due to quasiparticles can not only distinguish between full equilibrium and nonequilibrium, but also discriminate different nonequilibrium regimes, at least for large gap asymmetry. 

\subsection{Magnetic field tuning}
\begin{figure*}[htp]
    \centering
    \includegraphics[]{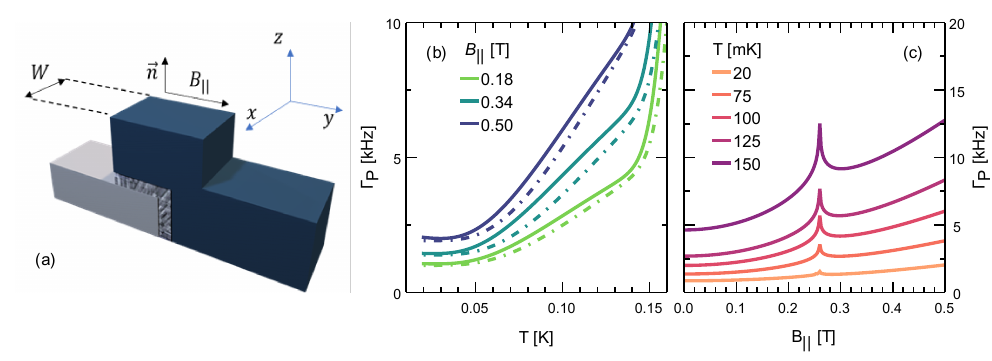}
 \caption{
    \textbf{Parity-switching rate in parallel magnetic field.}
    (a) Schematic of a Josephson junction with magnetic field $B_{||}$ applied in the plane of the junction, causing a phase gradient along the width $W$ of the junction. (b) Parity-switching rate vs temperature for different in plane fields. Solid curves are obtained using the full model, while the dot-dashed curves correspond to a global quasiequilibrium modeling with the same renormalized parameters used to capture the behavior of the zero-field curve (see caption to Fig.~\ref{fig:Fig4}). (c) Parity-switching rate vs in-plane magnetic field for different temperatures. The rate has a peak, at $B_{||}\simeq 0.27\,$T, when the resonant condition $\wq(B_{||})=\wLR$ is met.
    Parameters: $B_\Phi=0.8\,$T, $\wLR/(2\uppi)=5~$GHz, and the zero-field values of the remaining parameters are given in the caption of Fig.~\ref{fig:Fig3}.
    }
\label{fig:BparMultipanel}
\end{figure*}

While the concurrent measurement of excitation/relaxation and parity-switching rates has been demonstrated already~\cite{PRL121Exp}, it is more demanding than the determination of just the total parity-switching rate. Measuring the latter together with an accurate estimate of the gap asymmetry $\wLR$ would enable probing the various nonequilibrium regimes. Standard techniques for the measurement of $\wLR$, such as tunnel spectroscopy, are in general unavailable in qubits, and we estimate the uncertainty in $\wLR$ from quantum transport measurements on previously fabricated devices in the GHz range. In fact, nominally identical superconducting films displays gap variations of the order of a few up to tens of $\mu$eV~\cite{Court_2007}; at the same time, the determination of $\wLR$ via the so-called ``singularity-matching peak'' in the current-voltage characteristics~\cite{tinkham} has some intrinsic limitations, given its reduced visibility for $T\ll T_{\rm C}$ and the thermal broadening~\cite{PhysRevB.74.132501,Germanese2022}. As quasiparticle tunneling is enhanced near the resonant condition $\wq= \wLR$~\cite{MarchegianiQP}, tuning on-chip the qubit frequency $\wq$ is a viable strategy for estimating $\wLR$. The qubit frequency is commonly controlled by applying a magnetic field perpendicular to the plane of the junctions in a two-junction SQUID~\cite{OliverReview,BlaisRMP93}. In such a setup, it is possible to identify $\wLR$ by measuring the parity lifetime as a function of the perpendicular field; this strategy was successfully adopted in a recent experiment~\cite{Diamond22}, where a peak in the parity-switching rate was observed at the resonant condition. In particular, $\wLR$ was obtained with a resolution in the tens of MHz with this approach, which is suitable for $T\ll T_{\rm C}$.

An alternative way to tune the qubit frequency that can be used also in single-junction devices consists in applying a relatively strong magnetic field $B_{||}$, with typical values between tens to hundreds of mT for aluminum devices, in the plane of the junction, see schematic of Fig.~\ref{fig:BparMultipanel}a (the field is hence stronger than those usually applied in the perpendicular direction; nonetheless, transmon qubits have been shown to be resilient to in-plane fields, with lifetimes depending weakly on field up to a few hundred mT~\cite{Krause2022}).
The magnetic field modulates $\wq$ by means of two separate effects. First, the field weakens superconductivity in the two electrodes~\cite{tinkham}. The gaps of the two electrodes decrease monotonically with $B_{||}$, thus reducing $\wq$ (in a transmon, $\wq\propto \sqrt{E_{\rm J}}$, and the Josephson energy $E_{\rm J}$ is approximately proportional to the average gap); moreover, since the critical field for this modulation decreases with the film thickness (see, for instance, Ref.~\cite{Meservey1994}), 
the gap of the thicker film is typically suppressed more significantly, thus possibly affecting $\wLR$. Second, the penetration of the magnetic field through the junction oxide barrier makes the phase difference position dependent, varying along the direction $x$ in Fig.~\ref{fig:BparMultipanel}a, thus affecting directly the Josephson coupling. More precisely, the critical current of the junction is modulated by the magnetic field in a pattern resembling the Frauhnhofer diffraction by a narrow slit (see Appendix~\ref{app:Fraunhofer} for more details)~\cite{Barone}. The device geometry controls the relative size of these two suppression effects: for junctions with sufficiently thin electrodes, as the one experimentally investigated in field-resilient transmons~\cite{Krause2022,Krause24}, the frequency suppression is dominated by the Fraunhofer effect up to hundreds of milliTesla, with negligible gap-difference modulations; in practice, one could further reduce the field needed for frequency suppression and hence the field-induced change in $\wLR$ by properly choosing the aspect ratio of the junction (making it wider in the $x$ direction as compared to the $y$ one) while keeping fixed other parameters such as junction area and film thicknesses. For our goals, we assume for simplicity to be in such a regime, so to disregard any change in $\wLR$ due to the magnetic field $B_{||}$. 

Figure~\ref{fig:BparMultipanel}b displays the parity switching rate as a function of temperature for a few values of the in-plane magnetic field and large gap-asymmetry. The temperature dependence is qualitatively similar to the one already discussed above. Typically, the parity switching rates increase with the magnetic field; the exception is for fields such that $\wq(B_{||})\lesssim \wLR$, as discussed more extensively below. The solid curves are obtained using the full nonequilibrium model, treating the three quasiparticle densities as independent variables, while the dot-dashed curves are obtained assuming a global quasiequilibrium regime and renormalizing the parameters so as to match the zero field curve, see Fig.~\ref{fig:Fig4}a.
We note that at finite values of the magnetic field, the discrepancy between the full model and the global quasiequilibrium approach extends over a much wider temperature range compared to the zero-field case (cf. Fig.~\ref{fig:Fig4}a). These results suggest that it is not possible to use a consistent set of parameters to capture both the temperature dependence and the magnetic field dependence of the transition rates within the global quasiequilibrium approach. In other words, measuring the rates at different values of $B_{||}$ is a suitable approach to identify nonequilibrium regimes.

Moreover, the magnetic field evolution of the parity switching rates at a given temperature can give a direct probe of the gap asymmetry. In Fig.~\ref{fig:BparMultipanel}c, we display the parity switching rate as a function of the in-plane magnetic field for different temperatures; the dependence of the transition rates on $B_{||}$ is mainly related to interference effects due to the position-dependent phase difference across the junction (see Appendix~\ref{app:Fraunhofer}). The parity-switching rate evolution with field qualitatively resembles the behaviour observed in a SQUID transmon for perpendicular field up to half-flux quantum~\cite{Diamond22}; the rate typically increases with the field, except in the vicinity of $B_{||}\approx 0.27\,$T, where the peak signals again the resonance $\wq=\wLR$. As one can no longer ``renormalize'' the gap difference as to fit the data, the independent determination of $\wLR$ enables distinguishing between global and local quasiequilibrium, since the difference between the two approaches is manifest in the temperature regime for which $\mu_{\rm R>} = \mu_{\rm R<}\neq \mu_{\rm L}$ (cf. Fig.~\ref{fig:Fig3}b and Fig.~\ref{fig:BparMultipanel}b).
\section{Discussion}
\label{sec:conclusions}

We have developed a model based on rate equations to investigate the evolution with temperature of the qubit-quasiparticle system. Using effective chemical potentials, our model predicts different nonequilibrium regimes.
Superconducting qubits with gap asymmetry at most comparable to the thermal energy even at base temperature can largely be described with a single chemical potential, a situation which we call global quasiequilibrium. The competition between quasiparticle generation by pair-breaking photons and thermal phonons results in the commonly observed
crossover between the low-temperature limit, where the quasiparticles are out of equilibrium (finite chemical potential), and the high-temperature limit, where quasiparticles are thermalized to the substrate phonons (with vanishing chemical potential). 

In transmon with gap-asymmetric junctions, the quasiparticles in the two electrodes can be characterized by two different chemical potentials. This local quasiequilibrium regime can be observed for temperatures smaller than the gap asymmetry at which moreover quasiparticle-phonon scattering is fast enough to enable thermalization in the low-gap electrode. We discussed possible measurement strategies to discriminate the local and the global quasiequilibrium regimes, such as the joint measurement of excitation and relaxation parity-switching rates, or the estimate of these rates for different values of an in-plane magnetic field. Failing to account for the local nonequilibrium regime can result in inaccurate estimates of parameters such as the gap asymmetry and the photon-assisted generation rate. Conversely, the accurate determination of system parameters can be helpful in improving devices in which the precise targeting of parameter values is important, such as multi-qubit chips.

For temperatures at which tunneling is fast compared to quasiparticle-phonon scattering, our model predicts a third nonequilibrium regime with three different chemical potentials (and even a fourth regime with $\mu_{\rm L}-\mu_{\rm R>}\lesssim T\ll \mu_{\rm L} -\mu_{\rm R<}$ for small $\wLR$, cf. Appendix~\ref{app:lowwLRxQP}). However, such prediction is less reliable due to our thermalization ansatz [cf. Eqs.~\eqref{eq:fLAns} and \eqref{eq:fRAns}]. In fact, it is known that $f_{\rm L}(\e)$ and $f_{\rm R}(\e)$ can deviate from the Fermi distribution at low temperatures~\cite{Basko,Fischer2023,Fischer2024}; thus, a more accurate description of this regime requires determining the distribution function $f_{\rm L}(\e)$ and $f_{\rm R}(\e)$ as solutions of kinetic equations. The investigation of energy dependence of the quasiparticle distributions is beyond the scope of the present analysis; nevertheless, the estimates for the transition rates should be at least qualitatively correct, since the rates are mainly proportional to the quasiparticle densities (see Appendix~\ref{app:matTra}).
Indeed, our modeling has a weak dependence on the quasiparticle temperature $T$ for $T\ll \wLR,\,\w_{10}$, in agreement with previous work~\cite{MarchegianiQP}. We remark that our notion of quasiequilibrium differs from the one commonly used for charge and heat transport in normal-metal and superconducting junctions~\cite{Giazotto2006,Muhonen_2012}. In those experiments, the electron-electron scattering rate is fast compared to the electron-phonon one, and a thermal gradient between the quasiparticle and the phonon systems can be established. In superconductors, this regime can be 
realized if a large fraction of Cooper pairs is broken (that is, if the normalized densities $x_\alpha$ are not too small), as is generally the case for temperatures not too low compared to the critical one. Here we have considered parameter regimes in which
the quasiparticle densities and temperatures are low ($x_\alpha \ll 1$ and $T\ll T_{\rm C})$; then the quasiparticle-phonon scattering is typically faster than that due to quasiparticle-quasiparticle interaction~\cite{Reizer1989,Devereaux1989}, and so the latter can be ignored (cf. Appendix~\ref{app:qp-phonon}).

In this work we have neglected any mechanism leading to quasiparticle trapping. Quasiparticles can be trapped away from the junction through gap-engineering techniques, such as controlling the film thickness during deposition~\cite{Aumentado2004} or by adding normal-metal~\cite{RiwarPRB94} or superconducting traps~\cite{RiwarPRB100}. Quasiparticles can also be trapped in the core of vortices which can be present in a device because of
residual magnetic fields~\cite{Wang2014}. Phenomenologically, trapping can be incorporated in our rate equations by adding negative terms proportional to the densities to the right-hand sides of Eqs.~\eqref{eq:dotxL}-\eqref{eq:dotxR<}~\cite{MarchegianiQP}  
and could lead to additional nonequilibrium regimes in which the relations $\mu_{\rm L}\geq \mu_{\rm R>}\geq \mu_{\rm R<}$ may be violated. The proper inclusion of one or more trapping processes is left to future research.
\section{Methods}
The steady-state values of the quasiparticle densities have been obtained by setting to zero the left hand side of the rate equations Eq.~\eqref{eq:dotxL}-\eqref{eq:dotxR<}. The solution to this nonlinear algebraic system of equations has been found using standard findroot algorithms. The numerical results have been benchmarked with the analytical approximations derived in limiting cases.

\section{Data availability} 

The datasets used and analyzed in this paper are available from the corresponding author upon reasonable request.

\section{Code availability} 

The code used to generate the data for this paper is available from the corresponding author upon reasonable request.

\section{acknowledgments}

We acknowledge L. Amico, C. Dickel, and J. Krause for fruitful discussions. G.C. acknowledges support by the German Federal Ministry of Education and Research (BMBF), funding program ``Quantum technologies -- from basic research to market'', project QSolid (Grant No. 13N16149).

\section{Author contributions} 

G.M. developed the model, performed the numerical analysis, and drafted the manuscript. G.C. edited the manuscript, and provided supervision and guidance during the project. All authors contributed to the discussion and interpretation of the results.

\section{Competing interests} 

All authors declare no financial or non-financial competing interests.

\appendix
\section{Quasiparticle densities and chemical potentials}
\label{app:XqpandChemPot}
We consider here the relations between the normalized quasiparticle densities and the chemical potentials within the ansatz of Eqs.~\eqref{eq:fLAns} and~\eqref{eq:fRAns}. We treat $T/{\bar\Delta}$ and $\wLR/{\bar\Delta}$ [where $\DelBar=(\dL+\dR)/2$ is the average gap] as small parameters. These assumptions are appropriate in practice since, as temperature increases, the quasiparticle density and hence the qubit relaxation rate grows exponentially; similarly, a qubit with frequency comparable to the sum of the gaps could relax quickly by breaking Copper pairs. Then,
the superconducting density of states (DoS) can be approximated as follows~\cite{CatelaniPRB84}, 
\begin{equation}
\mathcal{N_\alpha}(\e)=\frac{\e}{\sqrt{\e^2-\Delta_\alpha^2}}\simeq \sqrt{\frac{\Delta_\alpha}{2(\e-\Delta_\alpha)}} \, ,
\label{eq:approxDoS}
\end{equation}
keeping the leading order term in the expansion around $\e=\Delta_\alpha$ ($\alpha={\rm L,R}$) of the non-singular part of the DoS [we recall that the superconducting DoS is gapped and so $\mathcal{N_\alpha}(\e)=0$ for $\e< \Delta_\alpha$]. Following the literature on quasiparticle decoherence~\cite{CatelaniPRL106,CatelaniSciPostReview}, the normalized quasiparticle density in the left electrode is defined as $x_{\rm L}=n_{\rm qp}^{\rm L}/n_{\rm cp}^{\rm L}$, dividing the quasiparticle density $n_{\rm qp}^{\rm L}$ by the Cooper pair density $n_{\rm cp}^{\rm L}=2\nu_0 \dL$; the expression reads 
\begin{equation}
\label{eq:xL}
\xL=\frac{2}{\dL}\int_{\dL}^{+\infty}d\e \,
\mathcal N_{\rm L}(\e) f_{\rm L}(\e)
\simeq
\sqrt{\frac{2\uppi T}{\Delta_{\rm L}}}
{\rm e}^{-(\dL-\mu_{\rm L})/T} \, .
\end{equation}
The approximation in Eq.~\eqref{eq:xL} is obtained using Eq.~\eqref{eq:approxDoS} and assuming a low temperature compared to the difference between the gap and the effective chemical potential  $T\ll \dL-\mu_{\rm L}$, so to approximate $f_{\rm L}(\e)\simeq {\rm e}^{-(\e-\mu_{\rm L})/T}$ in Eq.~\eqref{eq:xL}. We note that the chemical potential $\mu_{\rm L}$ (and so $\mu_{\rm R>}$ and $\mu_{\rm R<}$) is used for the quasiparticles, not for the electrons; in this sense, we do not assume any charge-mode disequilibrium. Equation~\eqref{eq:xL} is readily inverted to obtain the chemical potential as function of quasiparticle density and temperature,
\begin{equation}
    \mu_{\rm L} =\dL+T \log\left(x_{\rm L}\sqrt{\frac{\dL}{2\uppi T}}\right)\, .
\end{equation}

For the right electrode, similarly to Ref.~\cite{MarchegianiQP} we assume that, due to the competition between tunneling processes across the Josephson junction and quasiparticle-phonon scattering, the distribution function $f_{\rm R}$ takes the form in Eq.~\eqref{eq:fRAns}. Correspondingly, we can write the density $\xR$ as the sum of two contributions, $\xRm$ and $\xRp$, where 
\begin{align}
\label{eq:xRm}
\xRm&=\frac{2}{\dR}\int_{\dR}^{\dL}
d\e \,
\mathcal{N}_{\rm R}(\e)f_0(\e-\mu_{\rm R<})
\nonumber\\
&\simeq\sqrt{\frac{2\uppi T}{\dR}}
{\rm e}^{-(\dR-\mu_{\rm R<})/T}{\rm erf}[\sqrt{\wLR/T}]\, ,\\
\xRp&=\frac{2}{\dR}\int_{\dL}^{\infty}
d\e \,
\mathcal{N}_{\rm R}(\e)f_0(\e-\mu_{\rm R>})
\nonumber\\
&\simeq\sqrt{\frac{2\uppi T}{\dR}}{\rm e}^{-(\dR-\mu_{\rm R>})/T}\mathrm{erfc}[\sqrt{\wLR/T}]\, ,
\label{eq:xRp}
\end{align}
where $\textrm{erf}[z]$ is the error function and $\textrm{erfc}[z]=1-\textrm{erf}[z]$~\cite{abramowitz1968handbook}. As for $\xL$, 
the expressions in the second lines of Eqs.~\eqref{eq:xRm} and \eqref{eq:xRp} are obtained approximating $f_0(\e-\mu_{\rm R<(>)})$ with Boltzmann distributions, valid for $T \ll \dR-\mu_{\rm R<}\,, \,\dL - \mu_{\rm R>}$. We numerically verified for all our calculations that these conditions are satisfied over the considered range, thus justifying our approximation for the quasiparticle density. We note that for $\xRp$ a possible approximation for the DoS alternative to Eq.~\eqref{eq:approxDoS} is $\mathcal{N}_{\rm R}(\e)\simeq \dL/\sqrt{(\dL+\dR)(\e-\dR)}$, due to the lower limit of the integral (this approximation reproduces the one of Ref.~\cite{MarchegianiQP} further replacing $\e\to\dL$ in the limit $T\ll \wLR$). Since this expression is equivalent to Eq.~\eqref{eq:approxDoS} at the leading order in $\wLR/\DelBar$, we will use Eq.~\eqref{eq:approxDoS}, unless explicitly stated.

\subsection{Chemical potential differences and global quasiequilibrium}
\label{app:chemDiff}
Here, we summarize some bounds on the temperature dependence of the chemical potential differences, displayed in the inset of Fig.~\ref{fig:Fig3} in the main text. Combining Eq.~\eqref{eq:xRm} with Eqs.~\eqref{eq:xL} and~\eqref{eq:xRp}, we can immediately write: 
\begin{align}
\frac{\mu_{\rm L}-\mu_{\rm R<}}{\wLR}&= 1 + \frac{T}{\wLR} \log\left(\frac{\xL}{\xRm}\sqrt{\frac{\dL}{\dR}}{\rm erf}\sqrt{\wLR/T}\right) \, ,\\
\frac{\mu_{\rm R>}-\mu_{\rm R<}}{\wLR}&= \frac{T}{\wLR} \log\left(\frac{\xRp}{\xRm}\frac{{\rm erf}\sqrt{\wLR/T}}{{\rm erfc}\sqrt{\wLR/T}}\right)
\end{align}
For low temperatures compared to the gap difference, $T\ll \wLR$, at leading order we have ${\rm erf}[\sqrt{\wLR/T}]\simeq 1$, ${\rm erfc}[\sqrt{\wLR/T}]\simeq \sqrt{T/\uppi\wLR}\exp[-\wLR/T]$), and the chemical potential differences are larger (smaller) than $\wLR$ when $\xL$ and $\xRp\sqrt{\uppi\wLR/T}$ are larger (smaller) than $\xRm$, respectively. For large gap asymmetry, we have $\xRm\gg \xL,\xRp$~\cite{MarchegianiQP} (see also Appendix~\ref{app:largewLRxQP}),
making the logarithm argument smaller than one, and so the chemical differences smaller than $\wLR$ at low temperatures. Differently, for small gap asymmetry $\xL\gtrsim \xRp\sim \xRm$, making typically the chemical potential larger than $\wLR$ at temperature $T\sim \wLR$.

Note that in the local quasiequilibrium regime $\mu_{R>}=\mu_{R<}$ (cf. Fig.~\ref{fig:Fig1}b), the ratio between the quasiparticle densities in the low-gap electrode reads $\xRp/\xRm=\mathrm{erfc}[\sqrt{\wLR/T}]/\mathrm{erf}[\sqrt{\wLR/T}]$. If we further assume that the electrodes are in the global quasiequilibrium, the ratio between the densities in the two electrodes read $\xL/\xR=\sqrt{\dR/\dL}{\rm exp}[-\wLR/T]$. Using these relations, the steady state of the qubit-quasiparticle system in global quasiequilibrium can be computed using Eq.~\eqref{eq:Qubit_rate} and a single equation in the set  Eqs.~\eqref{eq:dotxL}- ~\eqref{eq:dotxR<}; practically, it is convenient to use the equation for the total quasiparticle density, obtained summing Eqs.~\eqref{eq:dotxL}-\eqref{eq:dotxR<} (see also Appendix~\ref{app:XqpandChemPot}).  

\section{Anharmonicity and rate equation for a transmon qubit}
\label{app:qubitPopulation}

In this section, we comment on two approximations made in modeling the superconducting qubit in the main text. First, we treated the superconducting qubit as an ideal two-level system. The low-energy spectrum of transmons is weakly anharmonic, so higher excited levels cannot generally be disregarded~\cite{BlaisRMP93,KehrerPRR6Qudit}, and can even be used as a resource for encoding~\cite{Elder2020,PRApplied17GHZ} and gate operations~\cite{DiCarlo2009,Fedorov2012}. For instance, treating the transmon as a qutrit~\cite{Qutrit_PRL126}, the population of the second excited state in thermal equilibrium is 
\begin{equation}
p_2=\frac{\exp[-(\omega_{10}+\omega_{21})/T]}{1+\exp(-\omega_{10}/T)+\exp[-(\omega_{10}+\omega_{21})/T]}\, .
\label{eq:p2Thermal}
\end{equation}
For a typical device with frequency $\w_{10}/(2\uppi)=5~$GHz and anharmonicity $(\w_{10}-\w_{21})/(2\uppi)\approx 300~$MHz, $p_2$ never exceed 10\% even for $T\approx 250~$mK, where the thermal energy is approximately equal to the transmon frequency. For this reason, in the main text we restrict our analysis only to the lowest two energy states, even though our results can be extended to incorporate the finite population of the second excited state. 

Second, in writing Eq.~\eqref{eq:Qubit_rate} we disregarded any asymmetry between the two parities of the transmon state. More generally, the rate equation for the population of the transmon state with logical value $i$ and parity $j$ reads~\cite{CatelaniPRB89,PRL121Exp}
\be
\dot{p}_i^j=-(\Gamma_{i\bar i}^{j\bar j}+\Gamma_{ii}^{j\bar j}+\Gamma_{i\bar i}^{jj})p_i^j
+\Gamma_{\bar i i}^{\bar j j}p_{\bar i}^{\bar j}
+\Gamma_{i i}^{\bar j j}p_{i}^{\bar j}
+\Gamma_{\bar i i}^{j j}p_{\bar i}^{j}\,
\label{eq:Qubit_rate_parity}
\ee
where $\bar i=i+1$ (mod 2), and $\bar {\rm e}={\rm o}$, $\bar {\rm o}={\rm e}$. For transmons, $E_{\rm J}/E_{\rm C}\gg 1$, the transition rates are approximately the same when exchanging even and odd parities, $\Gamma_{ii'}^{j\bar j}\simeq\Gamma_{ii'}^{\bar j j}$ and $\Gamma_{i i'}^{j j}\simeq\Gamma_{i i'}^{\bar j  \bar j}$ since the differences in the energies of even and odd states are small compared to temperature~\cite{CatelaniPRB89}. Adding the populations of the two parities, we arrive at the rate equation for the logical state occupation probability $p_i=p_{i}^{\rm e}+p_i^{\rm o}$ in Eq.~\eqref{eq:Qubit_rate}.

\section{Qubit transition rates due to quasiparticle tunneling}
\label{app:transitionRatesSingleJunction}
The transition rates for a superconducting qubit interacting with a quasiparticle bath can be computed with a standard perturbative approach using the golden-rule formula as in Ref.~\cite{CatelaniPRL106}. We denote the qubit states using a double index $|ij\rangle$, with $i\in\{0,1\}$ and $j\in\{\rm e,\,o\}$ accounting for the logical value and the parity, respectively. In particular, the transition between an initial state $|ij\rangle$ and a final state $|i'j'\rangle$ for a single-junction qubit reads 
\begin{align} \label{wif_gen}
\Gamma_{ii'}^{jj'} & = \left|\langle i'j'|\sin \frac{\hat\varphi}{2}|ij\rangle\right|^2
S^{+}_\qp\left(\w_{ii'}^{jj'}\right)
\\
& +\left|\langle i'j'|\cos \frac{\hat\varphi}{2}|ij\rangle\right|^2
 S^{-}_\qp\left(\w_{ii'}^{jj'}\right), \nonumber
\end{align}
where $\hat\varphi$ is the phase-difference operator across the Josephson junction, $S_\qp^\pm$ are the normalized quasiparticle
current spectral densities, and $\omega_{ii'}^{jj'}=\omega_i^{j}-\omega_{i'}^{j'}$ is the energy difference between the initial and the final states; we neglect from now on any parity-dependence of the qubit frequency, see Appendix~\ref{app:qubitPopulation}. Since each quasiparticle tunneling event changes the qubit parity, matrix elements between states with the same parity vanish for both $\sin\hat\varphi/2$ and $\cos\hat\varphi/2$.
\subsection{Temperature dependence of the quasiparticle current spectral densities}
\label{app:tempSqp}
We focus on qubits where the superconducting gaps and the quasiparticle distributions [cf. Eqs.~\eqref{eq:fLAns} and \eqref{eq:fRAns}] in the two electrodes are different. Thus, it is convenient to write the normalized quasiparticle
current spectral densities $S_\qp^\pm$ as the sum of the contributions from the two electrodes $S_\qp^\pm(\w)=\sum_{\alpha={\rm L,R}} S_{\alpha}^\pm(\w)$~\cite{MarchegianiQP}:  
\begin{align}
      S_{\alpha}^\pm(\w) =\frac{4g_{\rm T}}{e^2}\int_{\Delta_\alpha(\w)}^{+\infty}\!\!\! d\e \, 
\mathcal{N}_\alpha(\e)&\mathcal{N}_{\bar\alpha}(\e+\w)\nu_{\alpha\bar\alpha}^\pm(\e,\e+\w)
    \nonumber \\
    &\times f_\alpha(\e)[1-f_{\bar\alpha}(\e+\w)] \, ,
\label{eq:Sqp_general}
\end{align}
where $e$ is the absolute value of the electron charge, $g_{\rm T}$ is the normal-state conductance of the junction,  $\Delta_\alpha(\w)=\text{max}\{\Delta_\alpha,\Delta_{\bar\alpha}-\w\}$ and \begin{equation}
\nu_{\alpha\bar\alpha}^\pm(\e,\e') =\frac{1}{2}\left(1\pm \frac{\Delta_\alpha\Delta_{\bar\alpha}}{\e\e'}\right)
\end{equation} 
are the BCS coherence factors~\cite{tinkham}. In evaluating the spectral densities of Eq.~\eqref{eq:Sqp_general}, we consider the non-degenerate limit, where we can approximate $1-f_{\bar\alpha}\simeq 1$, consistently with our assumptions $T\ll \dL-\mu_{\rm L},\,\dL-\mu_{\rm R>},\,\dR-\mu_{\rm R<}$, see Appendix~\ref{app:XqpandChemPot}. The lower limit of integration in Eq.~\eqref{eq:Sqp_general} depends on the relation between the gap difference $\wLR$ and the frequency of the considered transition $\omega$. Keeping leading order contributions in $\wLR/\bar\Delta$ and $\omega/\bar\Delta$, the coherence factors approximately read
\begin{align}
\nu_{\alpha\bar\alpha}^+(\e,\e+\w)&\simeq 1\, ,
\label{eq:coherencePlusApprox}
\\
\nu_{\alpha\bar\alpha}^-(\e,\e+\w)&\simeq
\frac{2\e-\Delta_\alpha-\Delta_{\bar\alpha}+\w}{2\bar\Delta} \,.
\label{eq:coherenceMinusApprox}
\end{align}
Given our ansatz for the distribution function in the right electrode [see Eq.~\eqref{eq:fRAns}], the spectral density for $\alpha={\rm R}$ can be written as the sum of two contributions, $S^\pm_{\rm R}(\w)=S^\pm_{\rm R<}(\w)+S^\pm_{\rm R>}(\w)$, where $S^\pm_{\rm R>}(\w)$ accounts for quasiparticles with initial energy larger than $\dL$ tunneling from the right electrode. 

For $\w\leq 0$, that is for qubit excitation and parity-switching (within our approximations, $\omega_{ii}^{\rm eo}\simeq 0$) transitions, we have $S_{\rm R<}^\pm(\omega\leq 0)=0$; physically, this identity follows from the absence of available quasiparticle states for $\e<\dL$ in the left electrode. 
Using Eqs.~\eqref{eq:approxDoS} (approximating $\sqrt{\Delta_\alpha}\simeq \sqrt{\bar\Delta}$ in the numerator), \eqref{eq:coherencePlusApprox} and \eqref{eq:coherenceMinusApprox}, the spectral densities for $\alpha={\rm L}$ and (provided $\w\leq 0$) $\alpha={\rm R>}$ can be explicitly evaluated as 
\begin{align}
S_\alpha^{+}(\w)\simeq &\frac{2g_{\rm T}\bar\Delta}{e^2}{\rm e}^{-(\Delta_\alpha+\Delta_{\bar\alpha}-\w-2\mu_\alpha)/2T}\nonumber\\
&\times K_0\left[\frac{1}{2T}|\Delta_\alpha-\Delta_{\bar\alpha}+\w|\right] \, ,
\label{eq:SplusApp}
\\
S_\alpha^{-}(\w)\simeq &\frac{g_{\rm T}}{e^2}|\Delta_\alpha-\Delta_{\bar\alpha}+\w|{\rm e}^{-(\Delta_\alpha+\Delta_{\bar\alpha}-\w-2\mu_\alpha)/2T}\nonumber\\
&\times K_1\left[\frac{1}{2T}|\Delta_\alpha-\Delta_{\bar\alpha}+\w|\right] \, ,
\label{eq:SminusApp}
\end{align}
where $K_n(z)$ is the modified Bessel function of the second kind. With a slight abuse of notation, in Eqs.~\eqref{eq:SplusApp}-\eqref{eq:SminusApp} we have $\Delta_{\rm R>}\equiv\dR$. We stress that in the exponentials and in the arguments of the Bessel functions one cannot make approximations of the type $\Delta_\alpha \simeq \bar{\Delta}$, since the relevant energy scale to compare $\wLR$ to is $T$, not $\bar{\Delta}$. We note that our expressions include as a particular case (that of global quasiequilibrium, $\mu_\alpha \equiv \mu$) the Eqs.~(S12)-(S15) in the supplement to Ref.~\cite{connolly2023coexistence} upon the substitution ${\rm e}^{\mu/T}\to 1 + \zeta(T)x^\mathrm{res}_\mathrm{qp}/x_{\rm th}$, with $x_{\rm th} = \sqrt{2\uppi T/\Delta}e^{-\Delta/T}$. Note that the spectral densities are given in units of $2g_{\rm T}\Delta_\alpha/e^2$ in Ref.~\cite{connolly2023coexistence}, and due to the different notation, one should also replace $\dL\leftrightarrow\dR$, $\w \to - hf_{fi}$, and $\wLR \to \delta\Delta$.

Considering qubit relaxation processes, $\w> 0$, based on our ansatz of Eq.~\eqref{eq:fRAns} for $\alpha={\rm R<\,(R>)}$ the upper (lower) limit of the integral in Eq.~\eqref{eq:Sqp_general} is $\dL$. Consequently, the right-hand sides of Eqs.~\eqref{eq:SplusApp} and \eqref{eq:SminusApp} are modified for quasiparticles initially located in the right electrode. In particular, the modified Bessel functions of the second kind become incomplete 
\begin{align}
K_n(z)&\to K_n(z,w)  \quad\quad\quad\quad\quad
{\rm for} \quad \alpha={R>} \, ,\\
K_n(z)&\to K_n(z) - K_n(z,w)
\quad\, {\rm for} \quad \alpha={\rm R<} \, ,
\label{eq:replacement}
\end{align}
where $w=\cosh^{-1}[(\wLR+\w)/|\wLR-\w|]$, and
\begin{equation}
K_n(z,w)=\int_w^\infty \exp(-z \cosh t)\cosh[\nu t]\, ,
\label{eq:BesselIncompleteLow}
 \end{equation}
is the lower incomplete Bessel function~\cite{jones_2007}.

When the three chemical potentials are equal, i.e., the global  quasiequilibrium regime (cf. Fig.~\ref{fig:Fig1}c), the spectral densities obey the relation ($\bar {\rm L}= {\rm R}$ and $\bar {\rm R}= {\rm L}$)
\begin{equation}
S_\alpha^\pm(\omega)={\rm e}^{-\w/T}S_{\bar\alpha}^\pm(-\omega)\, ,
\label{eq:detailedBalanceSqp}
\end{equation}
which implies that the rates computed using Eq.~\eqref{wif_gen} satisfy the detailed balance principle in this regime. We note that the spectral densities and, consequently, the rates depend on the chemical potentials $\mu_\alpha$. Yet, by using the expressions Eqs.~\eqref{eq:xL}-\eqref{eq:xRp}, the dependence on the chemical potential can parametrized in terms of the quasiparticle densities $x_\alpha$, yielding the coefficients $\bar\Gamma_{ii'}^\alpha$ of the dynamical equations Eqs.~\eqref{eq:dotxL}-\eqref{eq:dotxR<} (in those equation, the rates are further normalized by the quasiparticle number in the right electrode).

\subsection{Next-to-leading order corrections to the spectral densities}
\label{app:NextToLeading}
Here, we consider next-to-leading order corrections to the spectral density functions computed above, where first-order corrections in $\omega/\bar\Delta$ and $\wLR/\bar\Delta$ are included; these expressions are computed by expanding the products of the densities of states and coherence factors as follows: 
\begin{align}
\label{eq:expansionNextDoStimescoherencePlus}
&\mathcal{N}_\alpha(\e)\mathcal{N}_{\bar\alpha}(\e+\w)\nu^+_{\alpha{\bar{\alpha}}}(\e,\e+\w)
\simeq \frac{\DelBar}{2\sqrt{\e-\Delta_\alpha}
\sqrt{\e-\Delta_{\bar{\alpha}}+\w}}\nonumber\\
&\times \left[1+\frac{2\e-(\Delta_\alpha+\Delta_{\bar\alpha}-\w)}{4\DelBar}\right]\\
&\mathcal{N}_\alpha(\e)\mathcal{N}_{\bar\alpha}(\e+\w)\nu^-_{\alpha{\bar{\alpha}}}(\e,\e+\w)
\simeq \frac{1}{4\sqrt{\e-\Delta_\alpha}
\sqrt{\e-\Delta_{\bar{\alpha}}+\w}}\nonumber\\
&\left[2\e-\Delta_\alpha-\Delta_{\bar\alpha}+\w + \frac{\Delta_\alpha-\Delta_{\bar \alpha}-\omega}{4\bar\Delta}(\Delta_\alpha-\Delta_{\bar \alpha}+\omega)\right]\, .
\label{eq:expansionNextDoStimescoherenceMinus}
\end{align}
When Eqs.~\eqref{eq:expansionNextDoStimescoherencePlus} and \eqref{eq:expansionNextDoStimescoherenceMinus} are plugged into Eq.~\eqref{eq:Sqp_general}, we find
\begin{align}
S_\alpha^{+}(\w)\simeq &\frac{2g_{\rm T}\bar\Delta}{e^2}{\rm e}^{-(\Delta_\alpha+\Delta_{\bar\alpha}-\w-2\mu_\alpha)/2 T}\nonumber\\
&\times \{K_0\left[\frac{|\Delta_\alpha-\Delta_{\bar\alpha}+\w|}{2 T}\right]
\nonumber\\
+ &\frac{|\Delta_\alpha-\Delta_{\bar \alpha}+\omega|}{4\bar\Delta} K_1\left[\frac{|\Delta_\alpha-\Delta_{\bar\alpha}+\w|}{2  T}\right]\} \, ,
\label{eq:SplusRef}
\\
S_\alpha^{-}(\w)\simeq &\frac{g_{\rm T}}{e^2}|\Delta_\alpha-\Delta_{\bar\alpha}+\w|{\rm e}^{-(\Delta_\alpha+\Delta_{\bar\alpha}-\w-2\mu_\alpha)/2 T}\nonumber\\
&\times \{ K_1\left[\frac{|\Delta_\alpha-\Delta_{\bar\alpha}+\w|}{2  T}\right] \, \nonumber\\
&+ \frac{\Delta_\alpha-\Delta_{\bar \alpha}-\omega}{4\bar\Delta}{\rm sign}(\Delta_\alpha-\Delta_{\bar \alpha}+\omega) \, \nonumber\\ &\times K_0\left[\frac{|\Delta_\alpha-\Delta_{\bar\alpha}+\w|}{2  T}\right]\} \,.
\label{eq:SminusRef}
\end{align}
For symmetric gaps ($\Delta_\alpha=\Delta_{\bar\alpha}$) these expressions reduce (up to the prefactor $2 g_{\rm T}/e^2$) to Eqs.~(35) and (40) of Ref.~\cite{CatelaniPRB89}. Also, the next-to-leading order expression of Eq.~\eqref{eq:SplusRef} has been used in calculating the transition rates in Ref.~\cite{Krause24} (see Appendix~G there).

\subsection{Transition rates for a single-junction transmon}
\label{app:matTra}

To arrive at explicit approximate expressions for the quasiparticle tunneling rates in a single-junction transmon, we need the matrix elements of the operators entering the tunneling Hamiltonian [see Eq.~\eqref{wif_gen} and Ref.~\cite{CatelaniPRB84}]. For convenience, we introduce the short-hand notations $s_{ii'}=|\langle ij | \sin\frac{\hat\varphi}{2} | i'\bar{j} \rangle|$, and $c_{ii'}=|\langle ij | \cos\frac{\hat\varphi}{2} | i'\bar{j} \rangle|$ ($\bar{\rm e}={\rm o}$ and $\bar{\rm o}={\rm e}$). In a single junction transmon, we can express the matrix elements between qubit states of opposite parity as~\cite{CatelaniPRB89}:
\begin{align}
\label{matSin10}
s_{10}&\simeq \left(\frac{E_{\rm C}}{8E_{\rm J}}\right)^{1/4},\\
\label{matSin00}
s_{ii}&\propto \left| \sin \left(2\uppi n_{\rm g}\right) \right| \frac{|\varepsilon_i|}{\omega_{\rm p}} ,
\\
\label{matCos00}
c_{ii}&\simeq 1 - \left(i+\frac12\right)\sqrt{\frac{E_{\rm C}}{8E_{\rm J}}} - \frac32 \left(i+\frac14 \right)\frac{E_{\rm C}}{8E_{\rm J}},\\
\label{matCos10}
c_{10} &\propto \left| \sin \left(2\uppi n_{\rm g}\right) \right| \frac{\sqrt{|\varepsilon_1\varepsilon_0}|}{\omega_{\rm p}} \, ,
\end{align}
where the approximations are valid for $E_{\rm J}\gg E_{\rm C}$. In the formulas above, $n_{\rm g}$ is the dimensionless offset charge of the transmon, and $\varepsilon_i$ is the charge dispersion of level $i$ (that is, the maximum energy difference between odd and even parities)~\cite{transmonPhysRevA76}:
\begin{equation}
    \varepsilon_i =(-1)^i E_{\rm C} \sqrt{\frac{2}{\uppi}}\frac{2^{2i+2}}{i!}\left(\frac{8E_{\rm J}}{E_{\rm C}}\right)^{(2i+3)/4}{\rm e}^{-\sqrt{8E_{\rm J}/E_{\rm C}}} \,.
\end{equation}
We note that at leading order in $E_{\rm J}/E_{\rm C} \gg 1$, the right-hand side of Eq.~\eqref{matCos00} is independent of the logical state and that, due to their exponential suppression, $s_{ii}$ [Eq.~\eqref{matSin00}] and $c_{10}$ [Eq.~\eqref{matCos10}] can be set to zero.

We start by considering the (scaled) parity-switching rates conserving the transmon logical state: for a single-junction transmon the matrix element of the $\sin\hat{ \varphi}/2$ operator is exponentially suppressed [see Eq.~\eqref{matSin00}], and therefore we have $\tilde\Gamma_{ii}^{\alpha}
\simeq c_{ii}^2 S_{\alpha}^-(0)/x_\alpha$. Combining Eq.~\eqref{eq:SminusApp} with the expressions of the quasiparticle densities, Eqs.~\eqref{eq:xL} and \eqref{eq:xRp}, we obtain
\begin{align}
&\tilde\Gamma_{ii}^{\rm L}
\simeq c_{ii}^2\frac{g_{\rm T}\dL}{e^2}
\sqrt{\frac{\dL-\dR}{2\dL}}
\sqrt{\frac{2y}{\uppi}}{\rm e}^y K_1[y] \, ,
\label{eq:GammaiiL}
\\
&\tilde\Gamma_{ii}^{\rm R>}\simeq c_{ii}^2\frac{g_{\rm T}\dR}{e^2}
\frac{\dL-\dR}{\sqrt{2T\dR/\uppi}}\frac{{\rm e}^{-y}K_1[y]}{\uppi\mathrm{erfc}[\sqrt{2y}]} \, ,
\label{eq:GammaiiRplus}
\end{align}
where $y=\wLR/2T$. We note that for $T\ll \wLR$ the rates $\tilde\Gamma_{ii}^{\rm L}$, $\tilde\Gamma_{ii}^{\rm R>}$ are in agreement with the low-temperature modeling of Ref.~\cite{MarchegianiQP} at the leading order in $\wLR/\DelBar$; the rate $\tilde\Gamma_{ii}^{\rm R>}$ is multiplied by an additional factor $\sqrt{\uppi}/2\approx 0.886$ compared to Ref.~\cite{MarchegianiQP}, as in the latter we assumed the distribution in the right electrode constant in a small window of width $T_\mathrm{qp}$ above $\dL$, while here it decreases exponentially with energy, see Eq.~\eqref{eq:fRAns}. Moreover, in the limit $\wLR\ll T$, we recover the results of Ref.~\cite{CatelaniSciPostReview}.

Similarly, for the computation of the (scaled) relaxation rates, we observe that the $\cos\hat{\varphi}/2$ operator matrix element is exponentially suppressed, giving $\tilde\Gamma_{10}^\alpha\simeq s_{10}^2 S^+_\alpha(\w_{10})/x_\alpha$; then we proceed as for the logical-state-conserving transitions to find:
\begin{align}
&\tilde\Gamma_{10}^{\rm L}
\simeq s_{10}^2\frac{g_{\rm T}\DelBar}{e^2}
\sqrt{\frac{2\dL}{\wLR+\w_{10}}}
\sqrt{\frac{2y_+}{\uppi}}{\rm e}^{y_+} K_0[y_+]\, ,
\label{eq:Gamma10LApp}
\\
&\tilde\Gamma_{10}^{\rm R>}\simeq s_{10}^2\frac{g_{\rm T}\DelBar}{e^2}
\sqrt{\frac{2\dR}{|\w_{10}-\wLR|}}\sqrt{\frac{2|y_-|}{\uppi}}\frac{{\rm e}^{y_-}K_0[|y_-|,w]}{\mathrm{erfc}[\sqrt{\wLR/T}]}\, ,
\label{eq:Gamma10RplusApp}
\\
&\tilde\Gamma_{10}^{\rm R<}\simeq s_{10}^2\frac{g_{\rm T}\DelBar}{e^2}
\sqrt{\frac{2\dR}{|\wLR-\w_{10}|}}\nonumber\\
&
\,\,\,\,\,\,\,\,\,\,\,\,\,\,\,\,
\times\sqrt{\frac{2|y_-|}{\uppi}}{\rm e}^{y_-}\frac{K_0[|y_-|]-K_0[|y_-|,w]}{{\rm erf}[\sqrt{\wLR/T}]}\, ,
\label{eq:Gamma10RminusApp}
\end{align}
where $y_\pm =(\w_{10}\pm\wLR)/2T$. In the limit $T\ll|\wLR-\w_{10}|$, the rates are the same to the ones reported in Ref.~\cite{MarchegianiQP} at the leading order in $(\wq\pm\wLR)/\DelBar$. In particular, both $\tilde\Gamma_{10}^{\rm L}$ and $\tilde\Gamma_{10}^{\rm R>}$ are finite; the rate $\tilde\Gamma_{10}^{\rm R<}$ is finite for $\w_{10}>\wLR$ ($y_->0$), and exponentially suppressed $\tilde\Gamma_{10}^{\rm R<}\propto {\rm e}^{-(\wLR-\wq)/T}$ for $\w_{10}<\wLR$ ($y_-<0$); these two situations correspond to case I and case II in the modeling of Ref.~\cite{MarchegianiQP}, respectively. 

Finally, we move to the excitation rates, $\tilde\Gamma_{01}^\alpha\simeq s_{10}^2 S^+_\alpha(-\w_{10})/x_\alpha$, and we find:
\begin{align}
&\tilde\Gamma_{01}^{\rm L}
\simeq s_{10}^2\frac{g_{\rm T}\DelBar}{e^2}
\sqrt{\frac{2\dL}{|\wLR-\wq|}}
\sqrt{\frac{2|y_-|}{\uppi}}{\rm e}^{-y_-} K_0[|y_-|] \, ,
\label{eq:Gamma01LApp}
\\
&\tilde\Gamma_{01}^{\rm R>}\simeq s_{10}^2\frac{g_{\rm T}\DelBar}{e^2}
\sqrt{\frac{2\dR}{\w_{10}+\wLR}}\sqrt{\frac{2y_+}{\uppi}}\frac{{\rm e}^{-y_+}K_0[y_+]}{\mathrm{erfc}[\sqrt{\wLR/T}]} \, .
\label{eq:Gamma01RplusApp}
\end{align}
The excitation rate from the larger gap superconductor is finite for $\w_{10}<\wLR$ (case II of Ref.~\cite{MarchegianiQP}) and exponentially suppressed $\tilde\Gamma_{01}^{\rm L}\propto {\rm e}^{-(\w_{10}-\wLR)/T}$ for $\w_{10}>\wLR$ (case I of Ref.~\cite{MarchegianiQP}).
We note that excitation rates from the right electrode are exponentially suppressed $\tilde\Gamma_{01}^{\rm R>}\propto {\rm e}^{-\wq/T}$ (so neglected in the modeling of Ref.~\cite{MarchegianiQP}). 
\subsection{Assignment of excitation rates for quasiparticles initially located in the higher-gap electrode}
\label{app:assignmentGamma01L}
Here, we derive the expression of the dimensionless factor $\xi$ in the rate equations, Eqs.~\eqref{eq:dotxR>}-\eqref{eq:dotxR<}. The transition rate $\tilde{\Gamma}_{01}^{\rm L}$ describes a process where quasiparticle tunnels from the left to the right electrode, exciting the qubit. The quasiparticle loses an energy $\wq$ to excite the qubit: in our modeling this quasiparticle increases either $\xRm$ or $\xRp$ if the final quasiparticle energy is smaller or larger than $\dL$, respectively. Thus, quasiparticle contributing to $\xRp$ have initial energy larger than $\dL+\w_{10}$; the calculation of the rates is performed according to the general expression of Eq.~\eqref{wif_gen}, replacing $\Delta_\alpha(\w)\to\dL+\wq$ in Eq.~\eqref{eq:Sqp_general}. In the non-degenerate limit $1-f_{\rm R}\simeq 1$, and using the approximations for the quasiparticle DoS [Eq.~\eqref{eq:approxDoS}], and the coherence factors [Eqs.~\eqref{eq:coherencePlusApprox} and \eqref{eq:coherenceMinusApprox}], we find for the fraction
\begin{align}\label{eq:xidef}
& \xi=\\ & \frac{s_{10}^2 K_0\left[\frac{|\wq-\wLR|}{2 T},w\right]+c_{10}^2 \frac{|\wLR-\w_{10}|}{2\bar{\Delta}}K_1\left[\frac{|\wq-\wLR|}{2 T},w\right]}{s_{10}^2 K_0\left[\frac{|\wq-\wLR|}{2 T}\right]+c_{10}^2 \frac{|\wLR-\w_{10}|}{2\bar{\Delta}} K_1\left[\frac{|\wq-\wLR|}{2 T}\right]}\, , \nonumber
\end{align}
where $w=\cosh^{-1}[(\w_{10}+\wLR)/|\w_{10}-\wLR|]$. This fraction is a function of the temperature, the qubit frequency, the gap difference and generally of the matrix elements between different matrix states. The dependence on the matrix elements drops for a single junction transmon qubit, as the $c_{10}$ term is exponentially suppressed for $E_{\rm J}\gg E_{\rm C}$.
In the limit of low temperatures $T\ll |\w_{10}-\wLR|$, we can use the asymptotic expansion of the incomplete Bessel function for large arguments $z\gg 1$~\cite{jones_2007} 
\begin{align}
K_n(z,w)\sim & \sqrt{\frac{\uppi}{2z}}{\rm e}^{-z}{\rm erfc}\left[(2z)^{1/2}\sinh\left(\frac{w}{2}\right)\right]
\nonumber\\
+&\frac{1}{z}\left[\frac{\cosh(n w)}{\sinh(w)}-\frac{1}{2\sinh(w/2)}\right]{\rm e}^{-z\cosh(w)}\,,
\label{eq:KincExp2}
\end{align}
which returns the standard asymptotic expression for the modified Bessel  
$K_n(z)\sim \sqrt{\uppi/(2z)}{\rm e}^{-z}$ for $w\to 0$ (independent of $n$). Using $z=|\wq-\wLR|/(2T)$ and $\cosh(w)=(\w_{10}+\wLR)/|\w_{10}-\wLR|$, at leading order for large $z$ (and finite $w$) we find 
\begin{equation}
 \frac{K_n(z,w)}{K_n(z)} \simeq \sqrt{\frac{T|\w_{10}-\wLR|}{\uppi \w_{10}\wLR}}\cosh(nw){\rm e}^{-\min\{\w_{10},\wLR\}/T}, 
\end{equation}
which gives rise to the exponential suppression of $\xi$ quoted in the main text. We note that the expansion in Eq.~\eqref{eq:KincExp2} can also be used to determine the low-temperature behavior of the scaled rates $\tilde\Gamma_{10}^{\rm R>}$ and $\tilde\Gamma_{10}^{\rm R<}$ in Eqs.~\eqref{eq:Gamma10RplusApp} and ~\eqref{eq:Gamma10RminusApp}.

\section{Quasiparticle-phonon interaction}
\label{app:qp-phonon}
In this work we have assumed that the quasiparticle distribution functions are a generalization of the Fermi-Dirac one, cf. Eqs.~\eqref{eq:fLAns} and \eqref{eq:fRAns}. More generally, the distribution function is determined by a kinetic equation of the form $df_\alpha(\e)/dt = \mathrm{St}\{f_\alpha,n_\alpha\}$, where the collision integral $\mathrm{St}$ accounts for electron-phonon interaction and is a functional of the quasiparticle and phonon ($n_\alpha$) distribution functions in electrode $\alpha$. The expressions for the collision integral in the superconducting state can be found, for instance, in Refs.~\onlinecite{KaplanPRB14,ChangScalapinoPRB15}.  Here we assume a non-degenerate quasiparticle occupation, $f_\alpha \ll 1$, a condition that validates neglecting the residual quasiparticle-quasiparticle interaction in the collision integral; in fact, the scattering and recombination rate due to electron-electron interaction scale with an additional power of $x_\alpha$ compared to the electron-phonon interaction, and so at the thermal equilibrium is exponentially suppressed for $T\ll T_{\rm C}$~\cite{Reizer1989,Devereaux1989}. This dependence physically expresses the necessity of having an extra quasiparticle that exchanges energy in a scattering process or absorbs the energy of the quasiparticle pair in a recombination process. Then the collision integral consists of four terms, $\mathrm{St} = \mathrm{St}_{\rm in} - \mathrm{St}_{\rm out} + \mathrm{St}_{\rm gen} - \mathrm{St}_{\rm rec}$: an incoming term $\mathrm{St}_{\rm in}$ due to relaxation of quasiparticle from higher energy via phonon emission or excitation from lower energy via phonon absorption, an additional incoming term $\mathrm{St}_{\rm gen}$ due to pair-breaking, an outgoing term $\mathrm{St}_{\rm out}$ due to relaxation to lower energy or excitation to higher energy, and a second outgoing term $\mathrm{St}_{\rm rec}$ due to the recombination of two quasiparticles into a Cooper pair. 

The functionals entering the kinetic equation are
\begin{align}
    \mathrm{St}_{\rm in} =& 
    4\uppi \int_0^\infty d\w F_\alpha(\w) 
    U^-_\alpha(\e,\e+\w) 
    [n_\alpha(\omega)+1]f_\alpha (\e+\w)\nonumber\\
    &+
    4\uppi \int_0^{\e-\Delta_\alpha} d\w F_\alpha(\w) U^-_\alpha(\e,\e-\w) 
    n_\alpha(\omega)f_\alpha (\e-\w)\, ,
    \label{eq:scattering-In}
    \\
    \mathrm{St}_{\rm out} =& 
    4\uppi \int_0^\infty d\w F_\alpha(\w) 
        U^-_\alpha(\e,\e+\w) 
    n_\alpha(\omega)f_\alpha (\e)
    \nonumber\\
    +&4\uppi \int_0^{\e-\Delta_\alpha}\!\!\!\! d\w F_\alpha(\w) 
    U^-_\alpha(\e,\e-\w) 
[n_\alpha(\omega)+1]f_\alpha (\e)\, ,
     \label{eq:scattering-Out}
    \\
    \mathrm{St}_{\rm rec} =& 
    4\uppi \int_{\e+\Delta_\alpha}^{\infty} \!\!\!\!\!\!  d\w F_\alpha(\w) 
   U^+_\alpha(\e,\w-\e) 
  [n_\alpha(\omega)+1] f_\alpha(\w-\e)f_\alpha(\e)\, ,
    \label{eq:Strec}
    \\
     \mathrm{St}_{\rm gen} =& 
     4\uppi \int_{\e+\Delta_\alpha}^{\infty} \!\!\!\!\!\!  d\w F_\alpha(\w) 
     U^+_\alpha(\e,\w-\e) 
     n_\alpha(\w)\, ,
     \label{eq:Stgen}
\end{align}
where we define the product of the DoS and the coherence factors as
\begin{equation}
U^\pm_\alpha(\e,\e')=\mathcal{N}_\alpha(\e')\nu^\pm_{\alpha\alpha}(\e,\e') \, ,
\label{eq:DoStimesBCSfactors}
\end{equation}
and the spectral function accounting for the matrix element of the electron-phonon interaction is
\be
F_\alpha(\w)=b_\alpha\w^2
\ee
with $b_\alpha$ a material-dependent parameter related to the electron-phonon coupling. The number of quasiparticles is conserved by the scattering terms ${\rm St}_{\rm in}$ and ${\rm St}_{\rm out}$: 
$\int_{\Delta_\alpha}^{\infty}\!d\e \, \mathcal{N}_\alpha(\e) \left[{\rm St}_{\rm in} - {\rm St}_{\rm out}\right]=0$ for arbitrary distribution functions.
Moreover, assuming equal quasiparticle and phonon temperatures (and $\mu_{R<}=\mu_{R>}$ for the right electrode), a Boltzmann distribution function for quasiparticles, $f_\alpha(\e)={\rm e}^{-(\e-\mu_\alpha)/T}$, and Bose-Einstein distribution for phonons, $n_\alpha(\w)=({\rm e}^{\w/T}-1)^{-1}$, are steady-state solutions of the scattering terms, ${\rm St}_{\rm in} - {\rm St}_{\rm out} = 0$, for arbitrary chemical potential. In contrast, in the absence of additional generation mechanisms, the generation and recombination terms would set the chemical potentials to zero (that is, full equilibrium).

\subsection{Quasiparticle relaxation in the lower-gap electrode}

Starting from the above collision integrals, here we derive the temperature dependence of the quasiparticle relaxation rate ($\tau_{\rm R}^{-1}$) in the lower gap (right) electrode. We are interested in a physical process where quasiparticles with initial energy $\e>\dL$ relax to states with energy $\e'=\e-\w<\dL$, emitting a phonon with energy $\w>\e-\dL$. This process is accounted for by the term in the second line of Eq.~\eqref{eq:scattering-Out}; specifically, the rate is computed by multiplying by the density of states, integrating for energies $\e>\dL$, and considering the lower limit $\e-\dL$ in the $\omega$ integration; 
\begin{align}
    \frac{1}{\tau_{\rm R}}\xRp=&\frac{4\uppi b_{\rm R}}{\dR} \int_{\dL}^{+\infty}d\e\,
    \mathcal{N}_{\rm R}(\e) f_{\rm R}(\e)\nonumber\\
    \times&\int_{\e-\dL}^{\e-\dR} d\w \, \w^2 \frac{\e(\e-\w)-\dR^2}{\e\sqrt{(\e-\w)^2-\dR^2}} [1+n_\alpha(\w)]\,.
    \label{eq:relaxV2}
\end{align}
Using the approximations of Eq.~\eqref{eq:approxDoS} and $f_{\rm R}(\e)={\rm e}^{-(\e-\mu_{\rm R>})/T}$, we perform the double substitution in the integrals of Eq.~\eqref{eq:relaxV2}
\begin{equation}
\begin{cases}
\e=\dR (1+x)\\
\e-\w=\dR (1+y).
\end{cases}
\end{equation}
Factorizing the quasiparticle density $\xRp$ of Eq.~\eqref{eq:xRp} in Eq.~\eqref{eq:relaxV2}, we identify the relaxation rate
\begin{equation}
\label{eq:tauRel}
\tau_{\rm R}^{-1}=2\uppi b_{\rm R}\dR^3\frac{\sqrt{\dR/\uppi T}}{{\rm erfc}[\sqrt{\wLR/T}]}\ \mathcal{I}\left(\frac{\dR}{T},\frac{\wLR}{\dR}\right),
\end{equation}
where we introduced the dimensionless double integral
\begin{equation}
\mathcal{I}(a,b)=\int_{b}^{\infty}dx\frac{e^{-ax}}{\sqrt{x}}
\int_{0}^{b}dy (y-x)^2\frac{xy+x+y}{\sqrt{y(y+2)}}\frac{1}{1-{\rm e}^{-a(x-y)}}
\end{equation}
In the low-temperature limit ($T\ll \wLR\leftrightarrow ab\gg 1$), one can drop the exponential factor in the $y$-integral of $\mathcal{I}(a,b)$, and the double integral can be evaluated explicitly. Further assuming $\wLR\ll\dR$, we can express the relaxation rate at the next-to next to leading order in $T/\wLR$ as 
\begin{equation}
\label{eq:tauRelLowT}
\tau_{\rm R}^{-1}\simeq2\uppi b_{\rm R} \dR^3 \frac{64\sqrt{2}}{105}\left(\frac{\wLR}{\dR}\right)^{7/2}
\left(1+\frac{7}{2}
\frac{T}{\wLR}+7\frac{T^2}{\wLR^2}\right)
\end{equation}
generalizing to small but finite $T/\wLR$ the expression given in Ref.~\cite{MarchegianiQP} (when considering the leading order in $\wLR/\Delta_R$). 
\subsection{Excitation by thermal phonons absorption}
Using a procedure similar to the one discussed in the previous subsection, we can compute the excitation term for quasiparticles with initial energy $\e\in [\dR,\dL]$, and final energy $\e'>\dL$, absorbing a phonon with minimal energy $\w=\dL-\e$. This process is included in the first integral of Eq.~\eqref{eq:scattering-Out}, where after multiplication by the density of states, and integration over $\e$, we find 
\begin{align}
\frac{1}{\tau_{\rm E}}\xRm =
\frac{4\uppi b_{\rm R}}{\dR}&\int_{\dR}^{\dL}d\e
\,
\mathcal{N}_{\rm R}(\e)f_{\rm R<}(\e)\nonumber\\
\times& \int_{\dL-\e}^{\infty}d\w 
\,
\w^2 \frac{\e(\e+\w)-\dR^2}{\e\sqrt{(\e+\w)^2-\dR^2}}n(\w)
\end{align}
Using again the approximation in Eq.~\eqref{eq:approxDoS} and $f_{\rm R<}(\e)={\rm e}^{-(\e-\mu_{\rm R<})/T}$, performing the substitutions 
\begin{equation}
\begin{cases}
\e=\dR (1+y)\\
\e+\w=\dR (1+x)
\end{cases} \, ,
\end{equation}
and dividing the resulting expression by the quasiparticle density $\xRm$ of Eq.~\eqref{eq:xRp},
we obtain for the excitation rate
\begin{equation}
\label{eq:tauExc}
\tau_{\rm E}^{-1}=2\uppi b_{\rm R}\dR^3\frac{\sqrt{\dR/\uppi T}}{{\rm erf}[\sqrt{\wLR/T}]}\ \mathcal{I}\left(\frac{\dR}{T},\frac{\wLR}{\dR}\right)
\, .
\end{equation}
In the limit $T\ll\wLR$, one can show that the excitation rate is exponentially suppressed, i.e., $\tau_{\rm E}^{-1}\propto {\rm e}^{-\wLR/T}$.

\subsection{Quasiparticle recombination}
\label{app:qpr}
Since in this work we consider approximate expressions for the rates at the leading order in $T/\Delta_\alpha\ll 1$ (cf. Appendix~\ref{app:XqpandChemPot}), in Eq.~\eqref{eq:Strec} we can approximate $F_\alpha(\omega)\simeq b_\alpha (2\Delta_\alpha)^2$, $1+n_\alpha(\w)\simeq 1$, and use Eq.~\eqref{eq:coherenceMinusApprox}; then, integration of Eq.~\eqref{eq:Strec} in $\omega$ gives immediately ${\rm St_{rec}}=\uppi b_\alpha (2\Delta_\alpha)^3 x_\alpha f_\alpha(\e)$. Multiplying by the DoS and integrating over $\e$, it is straightforwardly found that $r^{\rm L}=8\uppi b_{\rm L} \dL^3$, and that the recombination coefficients for quasiparticles residing in the low-gap electrode are equal at the leading order, $r^{\rm R>}\simeq r^{<>}\simeq r^{\rm R<}\simeq 8\uppi b_{\rm R} \DelBar^3$. This result is consistent with the low-temperature modeling of Ref.~\cite{MarchegianiQP} (see Appendix~B3 there), since the differences between these coefficients go to zero in the limit $\delta\to 1$. 

\subsection{Generation by thermal phonons}
\label{sec:thermalPhonons}
Consistently with the approximations done for the recombination rate and at the leading order in $T/\Delta_\alpha\ll 1$,  we approximate $F_\alpha(\omega)\simeq b_\alpha(2\Delta_\alpha)^2$, $n_\alpha(\w)\simeq {\rm e}^{-\w/T}$, and use Eqs.~\eqref{eq:approxDoS} and~\eqref{eq:coherencePlusApprox} for the approximated DoS and the coherence factor, respectively. Performing the $\w$ integration in Eq.~\eqref{eq:Stgen}, we find 
\begin{equation}
{\rm St}_{\rm gen}\simeq 
4\uppi b_\alpha \Delta_\alpha(2\Delta_\alpha)^2 \sqrt{\frac{\uppi T}{2\Delta_\alpha}}{\rm e}^{-(\e+\Delta_\alpha)/T}
\end{equation}
The generation rate $g^\mathrm{\rm pn}_{\rm L}$ in the higher gap electrode given in the main text is immediately obtained after multiplication by the DoS [again with the approximation in Eq.~\eqref{eq:approxDoS}] and integration over energy. The expressions for the thermal generation by phonons in the right electrode, $g^\mathrm{\rm pn}_{\rm R<}$ and $g^\mathrm{\rm pn}_{\rm R>}$, are computed similarly, by integrating over energies $\e\in[\dR,\dL]$ and $\e>\dL$,
respectively, as per our ansatz in Eq.~\eqref{eq:fRAns}.

\section{Photon-assisted tunneling in gap-asymmetric junctions}
\label{app:PAT}
In Ref.~\cite{PRL123Photon} it was shown that the qubit transition rate due to the absorption of pair-breaking photons with energy $\w_\nu$ takes a form similar to the one in Eq.~\eqref{wif_gen}. Specifically, the expressions differ for the spectral density of the process considered, according to the replacement: 
\begin{equation}
\label{eq:SphotonRepl}
  S_{\rm qp}^\pm (\w_{if})\to \Gamma_\nu \frac{g_{\rm T} \dL}{8 g_{\rm K}} S_{\rm ph}^\pm \left(\frac{\w_\nu+\w_{if}}{\dL};\frac{\dR}{\dL}\right)\, ,
\end{equation}
where $g_{\rm K}=e^2/(2\uppi)$ is the conductance quantum and the dimensionless factor $\Gamma_\nu$ is proportional to the qubit-field coupling strength and the average photon number~\cite{PRL123Photon}. The dimensionless photon spectral density reads~\cite{MarchegianiQP}
\begin{align}
& S_{\rm ph}^\pm(x;z)=\int\limits_1^{+\infty}dy\int\limits_z^{\infty} dy'\frac{y y'\pm z}{\sqrt{y^2-1}\sqrt{y'^2-z^2}}\delta(x-y-y')\, . 
\label{eq:Sph_pm}
\end{align}
Here, we provide explicit expressions for the spectral densities of the photon-assisted tunneling for arbitrary gap ratio $z$. Clearly, $S_{\rm ph}^\pm(x;z)=0$ for $x<1+z$, as there is no sufficient energy for Cooper-pair breaking. Performing the integration over $y'$ using the delta function, we obtain:
\begin{align}
& S_{\rm ph}^\pm(x;z)=\theta(x-1-z)\int\limits_1^{x-z} dy\frac{y(x-y)\pm z}{\sqrt{y^2-1}\sqrt{(x-y)^2-z^2}} 
\, \nonumber\\
& \ =\theta(x-1-z)\Bigg\{\sqrt{x^2-(z-1)^2}E\left[\sqrt{\frac{x^2-(z+1)^2}{x^2-(z-1)^2}}\right]
\nonumber\\
& \ -2z\frac{1\mp 1}{\sqrt{x^2-(z-1)^2}}K\left[\sqrt{\frac{x^2-(z+1)^2}{x^2-(z-1)^2}}\right]\Bigg\}
\label{eq:SphExplicit}
\end{align}
with $E$ and $K$ complete elliptic integrals of the second and first kind, respectively~\cite{PrudnikovBook}. This expression generalizes the results of Ref.~\cite{CatelaniSciPostReview} to asymmetric gaps and has already been reported in Ref.~\cite{Krause24}. The total qubit transition rate due to pair-breaking photons can be thus expressed as:
\begin{equation}
\Gamma^{\rm ph}=p_0 (\Gamma_{00}^{\rm ph}+\Gamma_{01}^{\rm ph}) + p_1 (\Gamma_{11}^{\rm ph}+\Gamma_{10}^{\rm ph}) 
\end{equation}
where $\Gamma_{if}^{\rm ph}$ are given by Eq.~\eqref{wif_gen} using the replacement in Eq.~\eqref{eq:SphotonRepl} and Eq.~\eqref{eq:SphExplicit}. 

In order to compute $g_{\rm R>}^{\rm ph}$ and $g_{\rm R<}^{\rm ph}$ (see Sec.~\ref{sec:QPgen}), 
we evaluate the fraction of quasiparticles generated in the right electrode with energy between $\dR$ and $\dL$ [limits of the integral from $z$ to $1$ in the integration over $y'$ of Eq.~\eqref{eq:Sph_pm}]and the fraction with energies larger than $\dL$ [limits of the integral from $1$ to $\infty$ in the integration over $y'$ of Eq.~\eqref{eq:Sph_pm}], so that $S_{\rm ph}^{\pm}(x;z)=S_{\rm ph}^{<,\pm}(x;z)+S_{\rm ph}^{>,\pm}(x;z)$. In particular, we have $S_{\rm ph}^{>,\pm}(x;z)=0$ for $x<2$. The spectral density $S_{\rm ph}^{>,\pm}(x;z)$ can be expressed as
\begin{align}
 S_{\rm ph}^{>,\pm}(x;z)=&\theta(x-2)\int\limits_1^{x-1} dy\frac{y(x-y)\pm z}{\sqrt{y^2-1}\sqrt{(x-y)^2-z^2}} 
\, \nonumber\\
 \ =&\Bigg\{\sqrt{x^2-(z-1)^2}E\left[\phi
,\sqrt{\frac{x^2-(z+1)^2}{x^2-(z-1)^2}}\right]
\nonumber\\
\ &- 2z\frac{1\mp 1}{\sqrt{x^2-(z-1)^2}}F\left[\phi,\sqrt{\frac{x^2-(z+1)^2}{x^2-(z-1)^2}}\right]
\nonumber\\
&-\sqrt{\frac{(x-2)(1-z^2)}{x}}
\Bigg\}\theta(x-2)\, ,
\end{align}
where $\phi=\sin^{-1}\sqrt{
(x-2)(x+1-z)/[x(x-1-z)]}$ is the angle in the incomplete elliptic integrals, and $F$ is the elliptic integral of the first kind ($F[\uppi/2,k]=K[k]$)].

\section{Approximate steady-state formulas for the quasiparticle densities}
\label{app:QPDensApprox}
Here, we derive some analytical approximations for the steady-state quasiparticle densities (and consequently the chemical potentials) and the qubit's excited state population. As in Ref.~\cite{MarchegianiQP}, we address separately the two limits of ``small'' (case I) and ``large'' (case II) gap asymmetry, further accounting for the temperature dependence of the quasiparticle rates. In the steady state, the quasiparticle generation is balanced by recombination, which [summing Eqs.~\eqref{eq:dotxL}-\eqref{eq:dotxR<}] results in the equation
\begin{equation}
\label{eq:balanceQP}
\frac{g^{\rm L}}{\delta}+g^{\rm R}=\frac{r^{\rm L}}{\delta} \xL^2 +r^{\rm R>} \xRp^2 +r^{\rm R<} \xRm^2 +2 r^{<>}\xRp\xRm\,,
\end{equation}
where $g^{\rm R}=g^{\rm R<}+g^{\rm R>}$. For notational convenience, we also introduce the total rate for a single quasiparticle to tunnel from the left or from the right electrode, 
\begin{align}
{\bar\Gamma}^{\rm L} & = p_0 (\bar\Gamma_{00}^{\rm L}+\bar\Gamma_{01}^{\rm L})+ p_1 (\bar\Gamma_{11}^{\rm L}+\bar\Gamma_{10}^{\rm L})\, ,\\
{\bar\Gamma}^{\rm R>} & = p_0 (\bar\Gamma_{00}^{\rm R>}+\bar\Gamma_{01}^{\rm R>})+ p_1 (\bar\Gamma_{11}^{\rm R>}+\bar\Gamma_{10}^{\rm R>}) \, .
\end{align}

\subsection{Small gap asymmetry}
\label{app:lowwLRxQP}
For $\w_{10}>\wLR$, quasiparticles can tunnel from both sides of the junction when absorbing energy from the qubit, so they populate both electrodes with densities that become approximately the same as the asymmetry gets comparatively smaller than $\w_{10}$~\cite{MarchegianiQP}. Using Eq.~\eqref{eq:dotxR<}, we can formally express the steady-state density $\xRm$ in terms of $\xL$, $\xRp$ by solving a quadratic equation,
\begin{widetext}
\begin{equation}
\xRm= \frac{\sqrt{(r^{<>}\xRp+\tau_{\rm E}^{-1}+p_1\bar{\Gamma}^{\rm R<}_{10})^2+4r^{\rm R<}[g^{\rm R<}+p_0\left(1-\xi\right)\bar\Gamma_{01}^{\rm L}  \xL+\tau_{\rm R}^{-1}\xRp]}-(r^{<>}\xRp+\tau_{\rm E}^{-1}+p_1\bar{\Gamma}^{\rm R<}_{10})}{2r^{\rm R<}}\,.
\label{eq:xrmsteady}
\end{equation}
\end{widetext}
To find approximate expressions for $\xL$ and $\xRp$, we first analyze the case $T\ll \wLR$, so that processes leading to the energy redistribution in the low-gap electrode are slow compared to tunneling. More precisely, quasiparticle excitation is exponentially suppressed with $\wLR/T$ and relaxation is proportional to $(\wLR/\dR)^{7/2}\ll 1$  (cf. Appendix~\ref{app:qp-phonon}); these terms can be ignored in Eq.~\eqref{eq:dotxR>} under the conditions $\bar{\Gamma}^{\rm L}-(1-\xi)\bar{\Gamma}_{01}^{\rm L}\gg \tau_{\rm E}^{-1}$ (in writing this condition, we use the fact that $\xL> \xRm$ for $T\ll\wLR\ll \w_{01}$) and $\bar{\Gamma}^{\rm R>}\gg \tau_{\rm R}^{-1}$. Furthermore, if the tunneling rates dominate over generation and recombination, $[\bar{\Gamma}^{\rm L}-(1-\xi)\bar{\Gamma}_{01}^{\rm L}]\xL\gg g^{\rm R>}$ and $\bar{\Gamma}^{\rm R>}\gg r^{\rm R>} \xRp+r^{<>} \xRm$, we find the following linear relation between $\xRp$ and $\xL$ using Eq.~\eqref{eq:dotxR>}, 
\begin{equation}
 \xRp=\frac{[\bar\Gamma^L -(1-\xi)\bar{\Gamma}_{01}^{\rm L}]\xL}{\bar\Gamma^{\rm R>}}
 \, .
\label{eq:balanceTunneling}
\end{equation}
The relation in Eq.~\eqref{eq:balanceTunneling} expresses the balance between the tunneling from the high-gap to low-gap electrode and \textit{vice versa} for quasiparticles with initial and final energies larger than $\dL$. The tunneling between the two electrodes is dominated by transitions preserving the qubit's logical state $\bar\Gamma^{\rm L} -(1-\xi)\bar{\Gamma}_{01}^{\rm L}\approx \bar{\Gamma}_{00}^{\rm L}$ and $\bar{\Gamma}^{\rm R>}\approx \bar{\Gamma}_{00}^{\rm R>}$ when the excited state population is not too large, i.e., $p_1\ll \bar{\Gamma}_{00}^{\rm L}/\bar{\Gamma}_{10}^{\rm L},\bar{\Gamma}_{00}^{\rm R>}/\bar{\Gamma}_{10}^{\rm R>}$~\cite{MarchegianiQP}; since the right-hand side in these inequalities is usually of order one and is increasing with $T$ (see the explicit formula in Appendix~\ref{app:matTra}), 
these conditions are typically satisfied for every temperature.
Thus, Eq.~\eqref{eq:balanceTunneling} implies that the chemical potentials for quasiparticles with energy larger than $\dL$ tends to equilibrate, i.e., $\mu_{\rm L}\simeq\mu_{\rm R>}$ while having at the same time $\mu_{\rm L}-\mu_{\rm R<}>T$;  this defines a fourth (nonequilibrium) regime as mentioned in Sec.~\ref{sec:conclusions}. This regime is practically hard to achieve for large $\wLR$ due to the increased relaxation rate ($\propto \wLR^{7/2}/\DelBar^{7/2}$) in the right electrode. Inserting Eq.~\eqref{eq:balanceTunneling} into Eq.~\eqref{eq:dotxL}, we find a quadratic equation for $\xL$ which depends on $\xRm$. However, when the population of the excited state is sufficiently small to have $p_1\bar{\Gamma}_{10}^{\rm R<}\xRm\ll g^{\rm L}/\delta$, we find 
\begin{equation}
    \label{eq:xL0sym}
\xL^{\rm I,(0)}=\frac{\sqrt{{[(1-\xi)\bar{\Gamma}_{01}^L]^2+4 g^{\rm L} r^{\rm L}}}-(1-\xi)\bar{\Gamma}_{01}^{\rm L}}{2r^{\rm L}}
\simeq\sqrt{\frac{g^L}{r^{\rm L}}}\, ,
\end{equation}
where the superscript ``${\rm I},(0)$'' denotes that these expressions are typically valid for $T\ll \wLR$ and gap asymmetry sufficiently smaller than $\w_{01}$. The approximation in Eq.~\eqref{eq:xL0sym} follows assuming $\bar\Gamma_{01}^{\rm L}\ll \sqrt{g^{\rm L} r^{\rm L}}$: this inequality generally holds for $T\ll \w_{10}-\wLR$ (so away from the resonant condition). In fact, the qubit excitation by a quasiparticle tunneling transition originating from the
left lead is exponentially suppressed
$\bar{\Gamma}_{01}^{\rm L}\propto\exp[-(\w_{10}-\wLR)/T]$ [see Eq.~\eqref{eq:Gamma01LApp} and text that follows], 
since there are no quasiparticle states available at energies $\e\approx \dL-\w_{10}<\dR$ in the right lead.  We note that Eqs.~\eqref{eq:balanceTunneling} and \eqref{eq:xL0sym} are equivalent to the results given in Sec.~IV A of Ref.~\cite{MarchegianiQP}, in the limits $\bar{\Gamma}^{\rm R>}\approx\bar{\Gamma}_{00}^{\rm R>}\gg  2\tau_{\rm R}^{-1},
4r^{\rm R>}\xRp $ and 
$[\bar{\Gamma}^{\rm L}-(1-\xi)\bar{\Gamma}_{01}^{L}]\xL\approx\bar{\Gamma}_{00}^{\rm L}\xL\gg g^{\rm R>}$
consistently with our assumptions. Nevertheless, we stress that here we do not assume the condition $\xRp\gg\xRm$ in deriving these expressions. The quasiparticle density $\xRm$ for $T\ll\wLR$ can be computed substituting Eqs.~\eqref{eq:balanceTunneling} and \eqref{eq:xL0sym} in Eq.~\eqref{eq:xrmsteady}, returning an expression which generalizes Eq.~(21) of Ref.~\cite{MarchegianiQP}, where we assumed $g^{\rm R<}=0$, $p_1=1-p_0=0$, and the rates $\tau_{\rm E}^{-1}$ and $\bar{\Gamma}_{01}^{\rm L}$ were neglected due to their exponential suppression for $T\ll \wLR$ and $T\ll\w_{10}-\wLR$, respectively. Using Eq.~\eqref{eq:xL0sym} (recall that $\xL>\xRm$), we can see that a sufficient condition for neglecting the term $ \delta p_1\bar{\Gamma}_{10}^{\rm R<}\xRm$ in Eq.~\eqref{eq:dotxL} is $p_1\ll \sqrt{g^{\rm L} r^{\rm L}}/\tilde{\Gamma}_{10}^{\rm R<}$, a condition which is well satisfied for our parameters choice and $T\lesssim \wLR$.

\begin{figure}[h]
  \begin{center}    \includegraphics[]{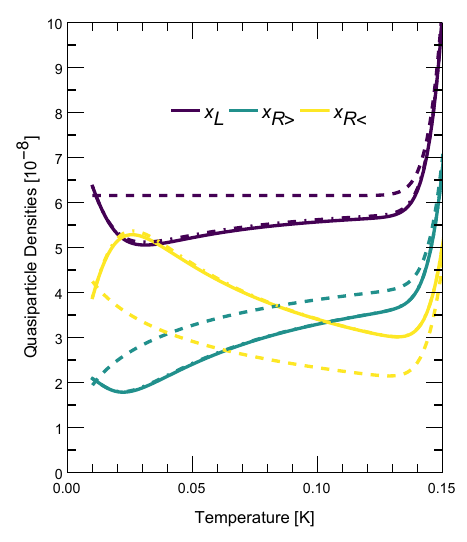}
  \end{center}
  \caption{
  \textbf{Quasiparticle densities vs temperature for small gap asymmetry.}
    Solid: numerical solution of Eqs.~\eqref{eq:Qubit_rate}-\eqref{eq:dotxR<} in the steady state, corresponding to the chemical potentials displayed in Fig.~\ref{fig:Fig3}a of the main text. Dashed: zeroth iteration formulas, Eqs.~\eqref{eq:balanceTunneling}-\eqref{eq:xRm0sym}. Dot-dashed: first iteration formulas, see Eqs.~\eqref{eq:xLSimCorr}-\eqref{eq:xRpSimCorr} and the text preceding them; we substituted in those formula  Eq.~\eqref{eq:p1Approx} for the excited state population of the qubit (where we can set $\tilde\Gamma_{01}^{\rm L}=\tilde\Gamma_{01}^{\rm R>}=0$).
  }
  \label{fig:FigQPSym}
\end{figure}
The above-mentioned approach only holds for $T\ll \wLR$, as the relaxation rate grows (polynomially) with $T/\wLR$ [see Eq.~\eqref{eq:tauRelLowT}] and gets quickly larger than $\bar{\Gamma}^{\rm R>}\approx \bar{\Gamma}_{00}^{\rm R>}$; similarly the excitation rate grows exponentially with $T/\wLR$ and becomes faster than $\bar{\Gamma}^{\rm L}\approx \bar{\Gamma}_{00}^{\rm L}$. 
However, the relation $\mu_{\rm L}\approx \mu_{\rm R>}$ is still valid for an arbitrary $T$, since the differences between the chemical potentials are typically smaller or comparable to $\wLR$ (see Appendix~\ref{app:chemDiff}), 
and so $\mu_{\rm L}-\mu_{\rm R>}<\mu_{\rm L}-\mu_{\rm R<}\ll T$ as the temperature grows and becomes comparable to $\wLR$ [cf. inset Fig.~\ref{fig:Fig3}a]. 
Thus, we construct an expression for the temperature dependence of the densities using an iterative procedure. Specifically, we write
$\xL\approx\xL^{\rm I,(0)}+\xL^{I,(1)}$ and
$\xRp\approx\xRp^{\rm I,(0)}+\xRp^{\rm I,(1)}$, where $\xL^{\rm I,(1)}\ll \xL^{\rm I,(0)}$ and $\xRp^{\rm I,(0)}$ is given by Eq.~\eqref{eq:balanceTunneling} replacing $\xL\to \xL^{\rm I,(0)}$. Even though we generally have $\xRp^{\rm I,(1)}<\xRp^{\rm I,(0)}$, this contribution is not necessarily small since the quasiparticle-phonon scattering rates (neglected in the derivation of $\xL^{\rm I,(0)}$ and $\xR^{\rm I,(0)}$) grow rapidly with $T$. If we neglect the contributions $\xL^{\rm I,(1)}$ and $\xRp^{\rm I,(1)}$ altogether, we can obtain an approximate expression for the temperature dependence of $\xRm$ using Eq.~\eqref{eq:balanceQP}, where, approximating $r^{\rm R<}\sim r^{\rm R>}\sim r^{<>}$ at leading order in $1-\delta\ll \delta$ (see Appendix~\ref{app:qpr}) and using Eq.~\eqref{eq:xL0sym}, we have 
\begin{equation}
\label{eq:xRm0sym}
\xRm^{\rm I,(0)}=\sqrt{g^{\rm R}/r^{\rm R<}}-\xRp^{\rm I,(0)} \, .
\end{equation}

To estimate the contributions $x_{\rm L}^{\rm I,(1)}$ and $x_{\rm R>}^{\rm I,(1)}$, we recall the quasiparticle generation by thermal phonons is negligible compared to photon pair-breaking for $T\leq\bar{T}$ [see Eq.~\eqref{eq:crossoverTemp}]. Generation due to the absorption of pair-breaking photons depends on the temperature only through the changing qubit's excited state population (see Appendix~\ref{app:PAT}); this dependence is weak as $\Gamma_{01}^{\rm ph}/\Gamma_{10}^{\rm ph}\sim 1$ and $\Gamma_{10}^{\rm ph}\ll \Gamma_{\rm 00}^{\rm ph}\sim\Gamma_{11}^{\rm ph}$ for a transmon with $E_{\rm J}\gg E_{\rm C}$. Thus, the total quasiparticle number is approximately fixed and, according to Eq.~\eqref{eq:balanceQP}, we can construct the ansatz $\xRm=\xRm^{\rm I,(0)}-(\xL^{\rm I,(1)}+\xRp^{\rm I,(1)})+\xRm^{\rm I,(2)}$, where $\xRm^{\rm I,(2)}\ll \xL^{\rm I,(1)}+\xRp^{\rm I,(1)}$. Neglecting quadratic terms in $\xL^{\rm I,(1)}$ and assuming that tunneling from the low-gap to the high-gap electrode is dominated by quasiparticles with $\e>\dL$, $\bar{\Gamma}^{\rm R>}\gg p_1\bar{\Gamma}_{10}^{\rm R<}$, we obtain using Eq.~\eqref{eq:dotxL} 
\begin{equation}
\label{eq:xLSimCorr}
\xL^{\rm I,(1)}=\frac{\bar{\Gamma}^{R>}\xRp^{\rm I,(1)}}{\bar{\Gamma}^{\rm L}+2r^L\xL^{\rm I,(0)}/\delta}\, ,
\end{equation}
which is equivalent to the expression in Ref.~\cite{MarchegianiQP} being $\bar{\Gamma}^{\rm L}\approx \bar{\Gamma}_{00}^{\rm L}$, $\bar{\Gamma}^{\rm R>}\approx \bar{\Gamma}_{00}^{\rm R>}$. 
By inserting Eq.~\eqref{eq:xLSimCorr} and the approximation $\xRm\approx\xRm^{\rm I,(0)}-(\xL^{\rm I,(1)}+\xRp^{\rm I,(1)})$ into Eq.~\eqref{eq:dotxR>}, we can find $\xRp^{\rm I,(1)}$ solving a quadratic equation (again approximating $r^{\rm R<}\sim r^{\rm R>}\sim r^{<>}$), namely
\begin{widetext}
\begin{align}
    \label{eq:xRpSimCorr}
    &\xRp^{\rm I,(1)}=
    [
    (\tau_{\rm E}^{-1} +\tau_{\rm R}^{-1}+\sqrt{g^{\rm R} r^{\rm R<}}) (\bar{\Gamma}^{\rm L} + 2 r^{\rm L} \xL^{\rm I,(0)}/\delta)
   + ( 2 r^{\rm L} \xL^{\rm I,(0)}/\delta +
    \tau_{\rm E}^{-1}-r^{\rm R<} \xRp^{\rm I,(0)}) \bar{\Gamma}^{\rm R>}
   \nonumber\\
   &- 
   \{[(\tau_{\rm E}^{-1} +\tau_{\rm R}^{-1}+\sqrt{r^{\rm R<} g^{\rm R}}) (\bar{\Gamma}^{\rm L} + 2 r^{\rm L} \xL^{\rm I,(0)}/\delta)
    - (2 r^{\rm L} \xL^{\rm I,(0)}/\delta +
    \tau_{\rm E}^{-1}-r^{\rm R<} \xRp^{\rm I,(0)}) \bar{\Gamma}^{\rm R>}]^2
      \nonumber\\
      &- 
    4 \bar{\Gamma}^{\rm R>} r^{\rm R<} (g^{\rm R>} 
    -\xRp^{\rm I,(0)}\sqrt{g^{\rm R} r^{\rm R<}} +  \tau_{\rm E}^{-1}\xRm^{\rm I,(0)} - \tau_{\rm R}^{-1}\xRp^{\rm I,(0)} )(
    \bar{\Gamma}^{\rm L} + 2 r^{\rm L} \xL^{\rm I,(0)}/\delta)\}^{1/2}]/(2 \bar{\Gamma}^{\rm R>} r^{\rm R<}) \, .
\end{align}
\end{widetext}

In Fig.~\ref{fig:FigQPSym} we compare the steady-state quasiparticle densities computed numerically by finding the roots of Eqs.~\eqref{eq:Qubit_rate}-\eqref{eq:dotxR<} (solid) and our analytical approximations. 
The zeroth iteration formulas in Eqs.~\eqref{eq:balanceTunneling}-\eqref{eq:xRm0sym} (dashed) are close to the numerical values at $T=10~$mK, but as temperature increases they generally overestimate $\xL$, $\xRp$ thus underestimating $\xRm$. Moreover, they don't capture the nonmonotonic behavior of the densities, since the quasiparticle relaxation rate $\tau_{\rm R}^{-1}$ in the right electrode is not accounted for. In contrast, the first iteration expressions in Eqs.~\eqref{eq:xLSimCorr}-\eqref{eq:xRpSimCorr} (dot-dashed) accurately describe the temperature dependence of the densities. More precisely, as $\tau_{\rm R}^{-1}$ grows with the temperature, $\xRp$ (and consequently $\xL$) decreases, causing an initial increase of $\xRm$ for temperatures $10~{\rm mK}\leq T\leq 25~{\rm mK}$; for larger temperatures, the thermal activation of the rate $\tau_{\rm E}^{-1}$ produces a reduction of $\xRm$, fixing the quasiparticle ratio in the low electrode to $\xRp/\xRm={\rm erfc}[\sqrt{\wLR/T}]/{\rm erf}[\sqrt{\wLR/T}]$ (see discussion in Appendix~\ref{app:chemDiff}), which grows monotonically with $T$. This behavior of the quasiparticle densities with temperature is reflected in the nonmonotonic dependence of the parity switching rate on $T$ (cf. Fig.~\ref{fig:Fig4}a).

\subsection{Large gap asymmetry}
\label{app:largewLRxQP}
\begin{figure}[bt]
  \begin{center}    \includegraphics[]{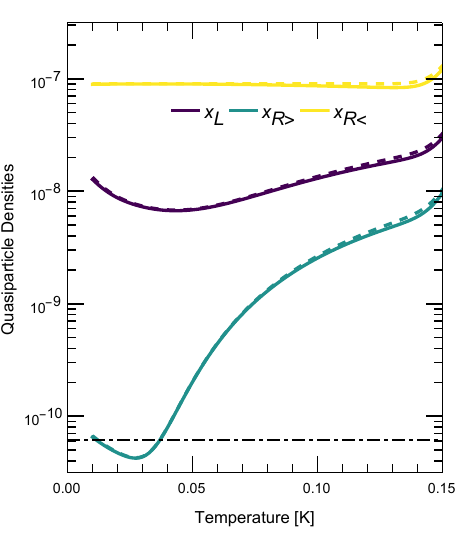}
  \end{center}
  \caption{
  \textbf{Quasiparticle densities vs temperature for large gap asymmetry.}
    Solid: numerical solution of Eqs.~\eqref{eq:Qubit_rate}-\eqref{eq:dotxR<} in the steady state, corresponding to the chemical potentials displayed in Fig.~\ref{fig:Fig3}b of the main text. Dashed: formulas in Eqs.~\eqref{eq:xrmss-II}-\eqref{eq:xrpss-II},
    with excited state population $p_1$ of Eq.~\eqref{eq:p1Approx} [neglecting $\tilde\Gamma_{01}^{\rm R>}\xRp\ll \tilde\Gamma_{01}^{\rm L}\xL$ and approximating $\xL\simeq (\Gamma_{00}^{\rm ph}+\Gamma_{00}^{\rm ph})/(\tilde\Gamma_{00}^{\rm L}+\tilde\Gamma_{01}^{\rm L})$ in this expression, see 
   Appendix~\ref{app:largewLRxQP}]. The horizontal dot-dashed line denotes the density corresponding to a single quasiparticle in the low-gap electrode.
  }
  \label{fig:FigQPAsym}
\end{figure}

Here, we focus on case II, with quasiparticles mainly located in the low-gap electrode, $\xRm\gg \xL\gg\xRp$, a condition typically obtained for a gap difference comparable to the qubit frequency due to fast energy relaxation in the low-gap electrode~\cite{MarchegianiQP}. 
We further assume that the generation at energies larger than the higher gap $\dL$ is dominated by pair-breaking photons, a condition valid for $T< \bar{T}$, with $\bar{T}$ the crossover temperature of Eq.~\eqref{eq:crossoverTemp}. For $\xL,\xRp\ll \xRm$, we can find an approximate expression for $\xRm$ using Eq.~\eqref{eq:balanceQP},
\begin{equation}
\label{eq:xrmss-II}
\xRm^{\rm II} =\sqrt{\frac{g^{\rm L}/\delta+g^{\rm R<}+g^{\rm R>}}{r^{\rm R<}}}\, .
\end{equation}
To compute the densities $\xL$ and $\xRp$, we proceed as in Ref.~\cite{MarchegianiQP}: when recombination terms can be ignored, $\delta\bar{\Gamma}^{\rm L}\gg r^{\rm L} \xL$ and $\bar{\Gamma}^{\rm R>}+\tau_{\rm R}^{-1}\gg r^{\rm R>} \xRp + r^{\rm R<} \xRm$, Eqs.~\eqref{eq:dotxL} and ~\eqref{eq:dotxR>} yield a linear system whose solution reads
\begin{widetext}
\begin{align}
 \xL^{\rm II} & =
     \,\frac{(g^{\rm L}/\delta+ p_1\bar{\Gamma}^{\rm R<}_{10} \xRm^{\rm II})(\bar\Gamma^{\rm R>}+ \tau_{\rm R}^{-1})+ (g^{\rm R>}+\tau_{\rm E}^{-1}\xRm^{\rm II})\bar\Gamma^{\rm R>}}{\bar\Gamma^{\rm L}\tau_{\rm R}^{-1}+\bar\Gamma^{\rm R>} p_0\bar\Gamma_{01}^{\rm L}(1-\xi)} \, ,\label{eq:xlss-II} \\
    \xRp^{\rm II} & =  \,\frac{(g^{\rm L}/\delta+ p_1\bar{\Gamma}^{\rm R<}_{10} \xRm^{\rm II})[\bar\Gamma^{\rm L}-p_0\bar\Gamma_{01}^{\rm L}(1-\xi)]+(g^{\rm R>}+\tau_{\rm E}^{-1}\xRm^{\rm II})\bar\Gamma^{\rm L} }{\bar\Gamma^{\rm L}\tau_{\rm R}^{-1}+\bar\Gamma^{\rm R>} p_0\bar\Gamma_{01}^{\rm L}(1-\xi)} \,.\label{eq:xrpss-II} 
\end{align}
\end{widetext}
The results obtained in Ref.~\cite{MarchegianiQP} are reproduced by setting $p_0=1-p_1=1$, assuming that all the quasiparticles are generated at energies larger than $\dL$, i.e., $g^{\rm R}=g^{\rm R>}=g^{\rm L}/\delta$, and for temperatures $T\ll \wLR,\, \wq$. In this limit, we can neglect quasiparticle excitation in the low-gap electrode through thermal phonon absorption, setting $\tau_{\rm E}^{-1}=0$ in Eqs.~\eqref{eq:dotxR>}-\eqref{eq:dotxR<}. Moreover, we can neglect the quasiparticle fraction at energies larger than $\dL$ for qubit's excitation processes: in other words we can disregard the terms proportional to $\bar\Gamma_{01}^{\rm R>}$ and set $\xi= 0$. 

In Fig.~\ref{fig:FigQPAsym} we compare the quasiparticle densities corresponding to the chemical potentials displayed in Fig.~\ref{fig:Fig3}b (solid) with the analytical expressions of Eqs.~\eqref{eq:xrmss-II}-\eqref{eq:xrpss-II} (dashed), finding good agreement over the temperature range up to $\bar{T}$. 
We indicate the density corresponding to a single quasiparticle in the right electrode with a horizontal dot-dashed line. For the parameters we use in our calculations, taking that line as a reference we see that both left and right electrodes host a large number of quasiparticles, hundreds to thousands at low temperatures; in this regime our approach, characterized by transition rates proportional to the quasiparticle densities [cf. Eqs.~\eqref{eq:Qubit_rate}-\eqref{eq:dotxR<}] is valid. When this condition is not satisfied, that is for densities such that $N_{\rm qp}\sim 1$ or less, quasiparticles cannot be simply considered as a memory-less bath for the qubit; for instance, if a single quasiparticle is present and absorbs energy from the qubit, the probability of a next tunneling event in general depends on the quasiparticle energy; moreover, fluctuations in the number of quasiparticles can lead to non-exponential decay of the qubit in time, see Refs.~\cite{Pop2014,QPpumping}.

\subsection{Excited state population of the qubit}
\label{sec:excitedQubit}
\begin{figure}
  \begin{center}    \includegraphics[]{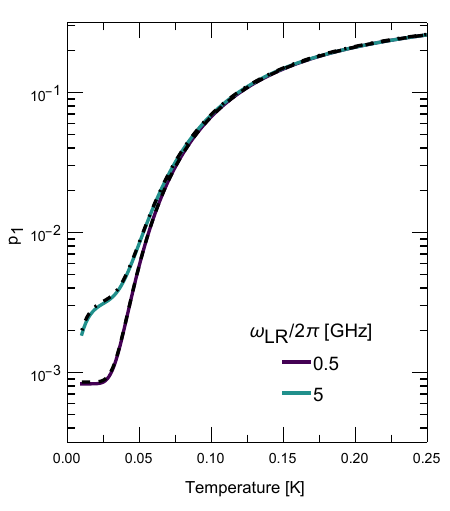}
  \end{center}
  \caption{
  \textbf{Excited state population of the qubit vs temperature.}
    Solid: numerical solution of Eqs.~\eqref{eq:Qubit_rate}-\eqref{eq:dotxR<} in the steady state, corresponding to the chemical potentials displayed in Fig.~\ref{fig:Fig3} of the main text. Dashed: Eq.~\eqref{eq:p1Approx} with $\xL=\xRp=0$. Dot-dashed: Eq.~\eqref{eq:p1Approx} with $\xRp=0$ and $\xL$ of Eq.~\eqref{eq:xlss-II} (setting $p_1=0$ in that equation, see Appendix~\ref{app:largewLRxQP}).
    }
  \label{fig:FigP1}
\end{figure}
The approximations derived in Appendices~\ref{app:lowwLRxQP} and \ref{app:largewLRxQP} are generally expressed in terms of the qubit's excited state population $p_1$, which depends self-consistently on the quasiparticle densities as follows 
\begin{equation}
p_1=\frac{\Gamma_{01}^{\rm ee}+\Gamma_{01}^{\rm ph}+\tilde{\Gamma}_{01}^{\rm L}\xL+\tilde{\Gamma}_{01}^{\rm R>}\xRp}{\Gamma_{10}^{\rm ee}+\Gamma_{01}^{\rm ee}+\Gamma_{10}^{\rm eo} +\Gamma_{01}^{\rm eo} 
} \, .
\label{eq:p1Expression}
\end{equation}
Typically, the lifetime $T_1=(\Gamma_{10}^{\rm ee}+\Gamma_{01}^{\rm ee}+\Gamma_{10}^{\rm eo} +\Gamma_{01}^{\rm eo})^{-1}$ of transmons is limited by processes not associated with quasiparticles~\cite{PRL121Exp}, for instance, losses due to spurious two-level systems~\cite{SiddiqiReview}. Hence, we have $\Gamma_{10}^{\rm ee}+\Gamma_{01}^{\rm ee}\gg\Gamma_{10}^{\rm eo},\Gamma_{01}^{\rm eo}$, and so the denominator of Eq.~\eqref{eq:p1Expression} is approximately independent of the quasiparticle-related processes. Differently, the qubit's excitation rate can be dominated by parity-switching contributions at low temperatures~\cite{PRL121Exp}; this is the case also in our calculations, since we assume the parity-conserving transition rates to respect the detailed balance principle, $\Gamma_{01}^{\rm ee}={\rm e}^{-\wq/T}\Gamma_{10}^{\rm ee}$, so that this excitation rate is exponentially small for $T\ll \wq$. Hence, we can approximate Eq.~\eqref{eq:p1Expression} as follows,
\begin{equation}
p_1=\frac{\Gamma_{01}^{\rm ph} +\tilde{\Gamma}_{01}^{\rm L}\xL+\tilde{\Gamma}_{01}^{\rm R>}\xRp}{\Gamma_{10}^{\rm ee}(1+{\rm e}^{-\wq/T})}+\frac{{\rm e}^{-\wq/T}}{1+{\rm e}^{-\wq/T}} \, .
\label{eq:p1Approx}
\end{equation}

In Fig.~\ref{fig:FigP1}, we display the temperature dependence of $p_1$ obtained by solving numerically the self-consistent system of Eqs.~\eqref{eq:Qubit_rate}-\eqref{eq:dotxR<} (solid curves) for the case of small and large gap asymmetry (cf. Fig.~\ref{fig:Fig3}). The excited state population is approximately the same for the two cases for temperatures $T\gtrsim 75~$mK, and its value follows the one dictated by the detailed balance principle, $p_1\simeq {\rm e}^{-\wq/T}/(1+{\rm e}^{-\wq/T})$. At lower temperatures, the excitation rate due to quasiparticle processes cannot be ignored, $p_1$ departs from thermal equilibrium expectations, and it generally depends on the gap asymmetry. 

For small gap asymmetry, the excitation rates via quasiparticle tunneling are exponentially suppressed at low temperatures, $\bar{\Gamma}_{01}^{\rm L}\propto {\rm e}^{-(\wq-\wLR)/T}$ and $\bar{\Gamma}_{01}^{\rm R>}\propto {\rm e}^{-\wq/T}$ for $T\ll \w_{10}-\wLR$ and $T\ll \w_{10}$ respectively [cf. Eqs.~\eqref{eq:Gamma01LApp}) and \eqref{eq:Gamma01RplusApp}]. Hence, ignoring the terms proportional to $\bar{\Gamma}_{01}^{\rm L}$ and $\bar{\Gamma}_{01}^{\rm R>}$ in Eq.~\eqref{eq:p1Approx}, the excited state population of the qubit is approximately independent of the quasiparticle densities. This approximate expression for $p_1$ agrees with our numerical findings, see Fig.~\ref{fig:FigP1}, where we show that $p_1$ is approximately independent of temperature up to a temperature of order $\wq/\ln (\Gamma_{10}^{\rm ee}/\Gamma_{01}^\textrm{ph}) \ll \wq$. 

For large gap asymmetry, $\wLR>\wq$, or close to resonance, $|\wLR-\wq|\sim T$, the rate $\tilde\Gamma_{01}^{\rm L}$ cannot be generally neglected. However, we can still ignore the term $\tilde{\Gamma}_{01}^{\rm R>}\xR$ in Eq.~\eqref{eq:p1Approx} due to the exponential suppression of $\tilde{\Gamma}_{01}^{\rm R>} \propto {\rm e}^{-\wq/T}$ combined with the fact that $\xRp\ll \xL$ for large gap asymmetry, see 
Appendix~\ref{app:largewLRxQP}.  
To find an approximate expression for $p_1$, we assume that the latter is sufficiently small that we can set
$p_1=0$ in Eq.~\eqref{eq:xlss-II} for $\xL$ (the conditions are $p_1\ll \bar\Gamma_{00}^{\rm L}/\bar\Gamma_{10}^{\rm L}$ and $ p_1\bar{\Gamma}^{\rm R<}_{10}\xRm^{\rm II}\ll g^{\rm L}/\delta$). Moreover, since typically for large gap asymmetry the relaxation rate is much faster than tunneling from the low-gap electrode, $\tau_{\rm R}^{-1}\gg \bar\Gamma^{\rm R>}$, Eq.~\eqref{eq:xlss-II} reduces to $\xL^{\rm II}\simeq (\Gamma_{00}^{\rm ph}+\Gamma_{01}^{\rm ph})/(\tilde{\Gamma}_{00}^{\rm L}+\tilde{\Gamma}_{01}^{\rm L})$. Using this approximation for $\xL$ in Eq.~\eqref{eq:p1Approx}, we find good agreement with the numerical results (see dot-dashed curve in Fig.~\ref{fig:FigP1}).

\subsection{Crossover temperature between nonequilibrium and local quasiequilibrium}
\label{app:crossoverNELocal}
Using the approximate expressions for the quasiparticle densities for large gap asymmetry, we derive an approximation for the crossover temperature between the nonequilibrium and the local quasiequilibrium regimes. As discussed in the main text, we identify this temperature equating the tunneling rate from the left to the right electrode at energies $\e\geq\dL$ and the quasiparticle excitation rate due to electron-phonon scattering,
\begin{equation}
\label{eq:balanceExcTun}
    [\bar{\Gamma}^{\rm L}-(1-\xi)\bar{\Gamma}_{01}^{\rm L}] \xL=\tau_{\rm E}^{-1}\xRm =\tau_{\rm R}^{-1}\frac{{\rm erfc}[\sqrt{\wLR/T}]}{{\rm erf}[\sqrt{\wLR/T}]}\xRm \,.
\end{equation}
The second equality above follows combining Eq.~\eqref{eq:tauRel} with Eq.~\eqref{eq:tauExc}. At the crossover we have $T\ll\omega_{\rm LR}$, so we can approximate ${\rm erf}[\sqrt{\wLR/T}]\simeq 1$, ${\rm erfc}[\sqrt{\wLR/T}]\simeq \sqrt{T/\uppi\wLR}\exp[-\wLR/T]$ and use Eqs.~\eqref{eq:tauRelLowT} and Eq.~\eqref{eq:xrmss-II} for the approximation of $\tau_{\rm R}^{-1}$ and $\xRm$ (in the latter, we neglect thermal generation by phonons). Moreover, using $x_{\rm L}^{\rm II}\approx (\Gamma_{00}^{\rm ph}+\Gamma_{01}^{\rm ph})/(\tilde{\Gamma}_{00}^{\rm L}+\tilde{\Gamma}_{01}^{\rm L})$ (see discussion at the end of Sec.~\ref{sec:excitedQubit}), the tunneling rate on the left-hand side of Eq.~\eqref{eq:balanceExcTun} is approximately equal to the photon generation rate in the left electrode $g^{\rm ph}_{\rm L}=\delta 
 g^{\rm ph}_{\rm R}$. Inserting these approximations in Eq.~\eqref{eq:balanceExcTun}, we finally arrive at the expression (keeping only the leading order term in $\wLR/\DelBar\ll 1$)
\begin{equation}
    \label{eq:TcrossNeLocal}
T^* = \frac{2\wLR}{W\left[ 2\cdot\frac{32^2}{105^2\uppi}\left(\frac{\wLR}{\DelBar}\right)^7\frac{r^{\rm R<}}{g_{\rm R}^{\rm ph}} \right]}   \, .
\end{equation}

\section{In-plane magnetic field effects on quasiparticle transitions}
\label{app:Fraunhofer}
We discuss here the effects of an external magnetic field applied in the plane of the Josephson junction (see schematic in Fig.~\ref{fig:BparMultipanel}a) on the transition rates. 
First, we address qubits with a single junction; for concreteness, we consider a rectangular junction area with length $l_1$ and width $l_2$ (denoted as $W$ in Fig.~\ref{fig:BparMultipanel}a). The superconducting phase difference across the JJ is generally position-dependent, according to the relation~\cite{Barone} 
\begin{equation}
    \nabla \varphi =\frac{2eL}{\hbar}\bm{{\rm B}}\times\bm{{\rm n}} \, ,
    \label{eq:phiVSh}
\end{equation}
where $L$ is the effective penetration depth of the magnetic field in the junction, $\bm{{\rm B}}$ is the total (external plus screening currents induced) in-plane magnetic field, and $\bm{{\rm n}}$ is the unit vector normal to the junction's plane. 
The effective length $L$ can be expressed as $L=\lambda_{\rm L} {\rm tanh}[d_{\rm L}/2\lambda_{\rm L}]+\lambda_{\rm R} {\rm tanh}[d_{\rm R}/2\lambda_{\rm R}]+d_{\rm AlOx}$, with $d_{\rm AlOx}$ thickness of the oxide layer; then $L\approx (d_{\rm L}+d_{\rm R})/2 +d_{\rm AlOx}$  when the London penetration depths ($\lambda_{\rm L},\,\lambda_{\rm R}$) are longer than the thicknesses ($d_{\rm L},\,d_{\rm R}$) of the two films~\cite{Barone,Weihnacht1969}.

Films used in the fabrication of superconducting qubits have a typical thickness in the tens of nanometers and junction sizes ranging from $\sim 100\,$nm to a few microns. Thus, the junction is  ``small''~\cite{Barone} with respect to the Josephson penetration length $\lambda_{\rm J}=\sqrt{\hbar/(2 e \mu_0 L J_1)}$, where $\mu_0=4\uppi\times 10^{-7}$A$\ $m$^{-1}$ is the magnetic permeability of free space, and $J_1\approx I_{\rm C}/(l_1 l_2)$ is the Josephson critical current density; for typical qubits with Josephson energies in the tens of GHz, using the Ambegaokar-Baratoff relation we can estimate $\lambda_{\rm J}$ to be at least several tens to a few hundreds of micrometers. Thus, the effect of screening currents is negligible, and the magnetic field is given by the external field $\bm{{\rm B}}\approx \Bpartwo \bm{{\rm \hat{x}}}+\Bparone \bm{{\rm \hat{y}}}$ [here $\bm{{\rm \hat{x}}}$ and $\bm{{\rm \hat{y}}}$ are the unit vectors in the transverse and longitudinal direction in the plane of the junction]. For a homogeneous magnetic field, the profile of the phase gradient is obtained by integration of Eq.~\eqref{eq:phiVSh}, giving  $\varphi(x)=\varphi+ 2eLB_{||,1}x/\hbar$ (for $B_{||,2}=0$), with the coordinate $x$ spanning the junction width, $x\in[-l_2/2, l_2/2]$.

Treating tunneling as taking place locally, we obtain the transition rates between the initial ($i$) and final ($f$) qubit states by averaging the position-dependent contributions over the junction width $l_2$
\begin{equation}
  \Gamma_{if}=\frac{1}{l_2}\int_{-l_2/2}^{l_2/2} \Gamma_{if}(x)dx \, ,
  \label{eq:gammaifFraunhofer}
\end{equation}
where the position-dependent rate is of the form [cf. Eq.~\eqref{wif_gen}]
\begin{align}
\label{eq:gammaifx}
\Gamma_{if}(x)& = \left|\braket{i|\sin\frac{\hat\varphi(x)}{2}|f}\right|^2 S^+(\w_{if})\\ & + \left|\braket{i|\cos\frac{\hat\varphi(x)}{2}|f}\right|^2 S^-(\w_{if}). \nonumber
\end{align}
In Eq.~\eqref{eq:gammaifx}, $\omega_{if}$ is the energy difference between the initial and the final qubit states, while the spectral densities $S^{\pm}$ are different for number-conserving quasiparticle tunneling events [see Eq.~\eqref{eq:Sqp_general}] and photon-assisted events [see Eq.~\eqref{eq:SphotonRepl}].
Using the trigonometric addition formula 
\begin{align}
\sin(\varphi + kx)&=\sin(\varphi)\cos(kx)+\cos(\varphi)\sin(kx)\\
\cos(\varphi + kx)&=\cos(\varphi)\cos(kx)-\sin(\varphi)\sin(kx) \, 
\end{align}
in Eq.~\eqref{eq:gammaifx}, the rate of Eq.~\eqref{eq:gammaifFraunhofer} can be written as
\begin{align}
\label{wif_gen_Fraun}
\Gamma_{if} = \left|\langle f|\sin \frac{\hat\varphi}{2}|i\rangle\right|^2
[
&(1+{\rm Sinc}[\uppi \Bparone/B_\Phi])S^{+}\left(\w_{if}\right)/2
\nonumber\\+&
(1-{\rm Sinc}[\uppi \Bparone/B_\Phi])S^{-}\left(\w_{if}\right)/2
]
\nonumber\\
+\left|\langle f|\cos \frac{\hat\varphi}{2}|i\rangle\right|^2
[
&(1-{\rm Sinc}[\uppi \Bparone/B_\Phi])S^{+}\left(\w_{if}\right)/2
\nonumber\\
+
&(1+{\rm Sinc}[\uppi \Bparone/B_\Phi])S^{-}\left(\w_{if}\right)/2].
\end{align}
with $B_\Phi=\Phi_0/L l_2$ the Fraunhofer field, $\Phi_0=h/2e$ the magnetic flux quantum, and ${\rm Sinc}[z]={\rm sin}[z]/z$. For $\Bparone=0$,  Eq.~\eqref{wif_gen_Fraun} takes the form of Eq.~\eqref{wif_gen}. 

\subsection{Single-junction transmon}
For a single junction transmon, the matrix element of the $\sin \hat{\varphi}/2$ operator is exponentially suppressed for transitions keeping fixed the transmon's logical state [c.f. Eq.~\eqref{matSin00}], while the matrix element of the $\cos \hat{\varphi}/2$ operator is exponentially suppressed for transitions changing the transmon's logical state [c.f. Eq.~\eqref{matCos10}]. Using Eq.~\eqref{wif_gen_Fraun} we can express the rates for logical-state conserving, relaxation, and excitation transitions as
\begin{align}
\Gamma_{ii}(\Bparone) &= c_{ii}^2
\left[
\frac{1-z}{2}S^+(\w_{eo})+
\frac{1+z}{2}S^-(\w_{eo})
\right] \, ,\\
\Gamma_{10}(\Bparone) &= s_{10}^2
\left[
\frac{1+z}{2}S^+(\w_{10})+
\frac{1-z}{2}S^-(\w_{10})
\right] \, \\
\Gamma_{01}(\Bparone) &= s_{10}^2
\left[
\frac{1+z}{2}S^+(-\w_{10})+
\frac{1-z}{2}S^-(-\w_{10})
\right]
\end{align}
where $z(\Bparone)=\text{Sinc}(\uppi \Bparone/B_{\Phi})$.
Since the excitation rate is modified compared to the zero-field case, we have to change accordingly the definition of $\xi$ in Eq.~\eqref{eq:xidef}.
Using the approach developed in this section, one can show that
\begin{equation}
\label{eq:xiMod}
\xi=\frac{(1+z)K_0\left[u,w\right]+ (1-z)\frac{|\wLR-\w_{10}|}{2\bar{\Delta}}K_1\left[u,w\right]}{ (1+z)K_0\left[u\right]+ (1-z)\frac{|\wLR-\w_{10}|}{2\bar{\Delta}}K_1\left[u\right]}
\end{equation}
with $u=|\wq-\wLR|/(2 T)$.

\subsection{Split transmon}
Here we consider a split transmon, in which the single junction is replaced by a SQUID comprising two Josephson junctions (here, and below, denoted with the index $n=\{\rm a,b\}$), as the one recently investigated in Ref.~\cite{Krause24}. The in-plane magnetic field modifies the quasiparticle rates in various ways. For an asymmetric SQUID, characterized by junctions with different geometrical dimensions as in Ref.~\cite{Krause24}, the Fraunhofer fields of the two JJs are different, i.e., $B_{\Phi \rm a}\neq B_{\Phi \rm b}$, being $B_{\Phi n}$ inversely proportional to the junction width $l_{2n}$. As a result, the dimensionless parameter expressing the asymmetry between the Josephson energy of the two junctions, i.e., $d=(E_{\rm Ja}-E_{\rm Jb})/(E_{\rm Ja}+E_{\rm Jb})$, is a function of the in-plane field. The Josephson energy of the $n$-junction reads 
\begin{equation}
\label{eq:Ejn}
E_{{\rm J}n}=\frac{g_n\Delta_L\Delta_R}{2\uppi g_{\rm K}(\dL+\dR)}K\!\left(\frac{|\dL-\dR|}{\dL+\dR}\right)\!\left|\text{Sinc}\left[
\uppi \frac{\Bparone}{B_{\Phi n}}
\right]\right|\!, 
\end{equation}
with $g_n$ normal-state conductance of the junction and $K[z]$ the complete elliptic integral of the first kind~\cite{PrudnikovBook}. The quantity in the absolute value of Eq.~\eqref{eq:Ejn} accounts for the Fraunhofer modulation of the critical current~\cite{Barone,Krause2022}, while the prefactor incorporates the gap-dependence of the Josephson current for different gaps $\dL\neq\dR$~\cite{Averin2021, Barone}. For $\Bparone=0$ and $\dL=\dR$, Eq.~\eqref{eq:Ejn} corresponds to the Ambegaokar-Baratoff relation for the critical current $I_{c,n}=2\uppi E_{Jn}/\Phi_0$~\cite{tinkham}. Using Eq.~\eqref{eq:Ejn}, the field-dependent asymmetry parameters $d(\Bparone)$ read
\begin{equation}
d(\Bparone)\equiv \frac{E_{\rm Ja}-E_{\rm Jb}}{E_{\rm Ja}+E_{\rm Jb}}=\frac{g_{\rm a} |z_{\rm a}(\Bparone)| - g_{\rm b} |z_{\rm b}(\Bparone)|}{g_{\rm a} |z_{\rm a}(\Bparone)| + g_{\rm b} |z_{\rm b}(\Bparone)|} \,,
\end{equation}
where $z_n(\Bparone)=\text{Sinc}(\uppi \Bparone/B_{\Phi n})$. 
The position dependence of the phase difference $\varphi(x)$ in an in-plane magnetic field modifies the transition rates as in Eq.~\eqref{wif_gen_Fraun}; in a split transmon, the total transition rate is the sum of the contributions of each junction~\cite{CatelaniPRB84, MarchegianiQP}
\begin{align}
\label{wif_split_transmon}
& \Gamma_{if} = \sum_{n={\rm a,b}} g_n\ \\ & \bigg\{ \left|\langle f|\sin \frac{\hat\varphi_n}{2}|i\rangle\right|^2
\left[
\frac{1+z_n}{2}\mathcal{S}^{+}\left(\w_{if}\right)+
\frac{1-z_n}{2}\mathcal{S}^{-}\left(\w_{if}\right)
\right]
\nonumber\\
&+\left|\langle f|\cos \frac{\hat\varphi_n}{2}|i\rangle\right|^2
\left[
\frac{1-z_n}{2}\mathcal{S}^{+}\left(\w_{if}\right)+
\frac{1+z_n}{2}\mathcal{S}^{-}\left(\w_{if}\right)
\right]\!\bigg\}. \nonumber
\end{align}
with $\mathcal{S}^\pm(\omega)=S_n(\omega)/g_n$ independent of the junction considered [cf. Eqs.~\eqref{eq:Sqp_general} and \eqref{eq:SphotonRepl}].
For a generic value of the reduced flux $f=\Phi/\Phi_0$ (where $\Phi$ is the magnetic flux threading the SQUID), the relevant matrix elements for the transitions are at the leading order in $E_{\rm J}(f)=E_{\rm Ja}+E_{\rm Jb}\gg E_{\rm C}$~\cite{CatelaniPRB84,MarchegianiQP},  
\begin{align}
\label{matSin10Split}
\left|\langle 1j | \sin\frac{\hat\varphi_n}{2} | 0\bar{j} \rangle\right|^2 &\simeq \left[\frac{E_{\rm C}}{8E_{\rm J}(f)}\right]^{1/2}\frac{1+\cos(\uppi f\pm\vartheta)}{2}\, ,\\
\label{matSin00Split}
 \left|\langle ij| \sin\frac{\hat\varphi_n}{2} | i\bar{j} \rangle\right|^2  &\simeq \frac{1-\cos(\uppi f\pm\vartheta)}{2}\, , \\
\label{matCos00Split}
\left|\langle ij | \cos\frac{\hat\varphi_n}{2} | i\bar{j} \rangle\right|^2 &\simeq \frac{1+\cos(\uppi f\pm\vartheta)}{2}\, ,\\
\label{matCos10Split}
\left|\langle 1j | \cos\frac{\hat\varphi_n}{2} | 0\bar{j} \rangle\right|^2  &\simeq\left[\frac{E_{\rm C}}{8E_{\rm J}(f)}\right]^{1/2}\frac{1-\cos(\uppi f\pm\vartheta)}{2} \, ,
\end{align}
where in the right hand side of these equations, $+$ is for $n={\rm b}$ and $-$ for $n={\rm a}$, and the angle $\vartheta$ satisfies the relation $\tan(\vartheta)=d \tan(\uppi f)$. Inserting Eqs.~\eqref{matSin10Split}-~\eqref{matCos10Split} into Eq.~\eqref{wif_split_transmon}, rates for transitions changing and preserving the qubit's logical state read respectively
\begin{align}
\label{eq:Gamma10QP}
\Gamma_{i\bar i} &=
\sqrt\frac{E_{\rm C}}{8E_{\rm J}(f)}
[
\gamma_+S_{\Sigma}^+(\w_{i\bar i})+
\gamma_-S_{\Sigma}^-(\w_{i\bar i})
]\\
\label{eq:GammaiiQP}
\Gamma_{ii} &=
[
\gamma_+ S_{\Sigma}^-(\w_{ii})+
\gamma_-S_{\Sigma}^+(\w_{ii})
]
\end{align}
where $S^\pm_\Sigma=(g_{\rm a}+g_{\rm b}) \mathcal{S}^\pm$ and we defined the weights 
\begin{equation}
\gamma_\pm = \frac{1}{2}\pm\frac{(z_+ + d_{0} z_-)\cos(\uppi f)^2+d(z_- + d_{0} z_+)\sin(\uppi f)^2}{4\mathcal{G}(f,d)}
\label{eq:weightsFraunplusSQUID}
\end{equation}
where $z_\pm =z_{\rm a}\pm z_{\rm b}$, $d_0=(g_{\rm a}-g_{\rm b})/(g_{\rm a}+g_{\rm b})$ is the asymmetry parameter for zero in-plane field, and $\mathcal{G}(f,d)$ is the Josephson energy suppression due to the flux through the SQUID, $\mathcal{G}(f,d)=\sqrt{\cos^2(\uppi f)+d^2\sin^2(\uppi f)}$. Finally, the parameter $\xi$ should be defined analogously to Eq.~\eqref{eq:xiMod} with the replacements $1\pm z \to \gamma_\pm$.

\end{document}